\documentclass[11pt,onecolumn,english,amsart,preprintnumbers,amsmath,amssymb,floatfix,nofootinbib]{revtex4}
\usepackage{tikz,xcolor}
\usepackage[colorlinks = true,  
linkcolor = red,
urlcolor  = blue,
citecolor = red,
anchorcolor = blue]{hyperref}   

\definecolor{lime}{HTML}{A6CE39}
\DeclareRobustCommand{\orcidicon}{%
	\begin{tikzpicture}
	\draw[lime, fill=lime] (0,0) 
	circle [radius=0.16] 
	node[white] {{\fontfamily{qag}\selectfont \tiny ID}};  
	\draw[white, fill=white] (-0.0625,0.095) 
	circle [radius=0.007];
	\end{tikzpicture}
	\hspace{-2mm}
}

\foreach \x in {A, ..., Z}{%
	\expandafter\xdef\csname orcid\x\endcsname{\noexpand\href{https://orcid.org/\csname orcidauthor\x\endcsname}{\noexpand\orcidicon}}
}

\usepackage[toc,page]{appendix}
\usepackage{epsfig}
\usepackage{graphics}
\usepackage{latexsym}
\usepackage{amsmath}
\usepackage{amssymb}
\usepackage{rotating}
\usepackage{subfigure}
\usepackage{bm}
\usepackage{color}
\usepackage{ulem}
\usepackage{booktabs}

\begin{document}


\title {  Self- similar behavior  of the Neutron Fracture Functions  }

\author{Samira Shoeibi Mohsenabadi$^{1}$\orcidA{}}
\email{Samira.Shoeibimohsenabadi@mail.um.ac.ir}

\author{F. Taghavi-Shahri$^{1}$\orcidC{}}
\email{Taghavishahri@um.ac.ir (Corresponing author)}

\affiliation {
$^{(1)}$Department of Physics, Ferdowsi University of Mashhad, P.O.Box 1436, Mashhad, Iran               }

\date{\today}

%
%
\begin{abstract}\label{abstract}
In this article, we have employed  fractal formalism to calculate the Fracture Functions of the Leading 
neutron produced in \textit{ep} collisions. The fractal concept describes the self-similar behavior
 of the proton structure at Leading neutron production of semi-inclusive Deep Inelastic Scattering at 
 the low values of the fractional momentum variable $\beta$. The Fracture Functions (FFs) parameterized 
 the non-perturbative part of the fragmentation process at the initial scale of  Q$_{0}^{2}$. In this analysis, we benefit 
 from the Leading neutron (Ln) experimental data published by H1 Collaboration and in order to estimate the uncertainty of neutron FFs and corresponding observables, we used the Hessian method. As a consequence of this Next-to-Leading 
 order  QCD analysis, we achieve a nice agreement between the prediction of our model for neutron FFs 
 and experimental data. It seems that the fractal approach based on the self-similar behavior of   these conditional parton distribution functions at low x can nicely describe the experimental data. 

\end{abstract}

\pacs{12.38.Bx, 12.39.-x, 14.65.Bt}

\maketitle

\tableofcontents{}

%
\section{Introduction}\label{sec:introduction}
Proton is known as a strongly bound state of quarks and gluons in Quantum Chromodynamics (QCD). Considering its collisions with electron not only clarify its structure, but also reveal the dynamics of strong interaction at high energy physics~\cite{Bourrely:2018qit,AbdulKhalek:2019bux,Zaidi:2019mfd,Harland-Lang:2014zoa,Ball:2014uwa,Martin:2009iq,Alekhin:2018pai,Salajegheh:2019zos}. Investigation of the Parton Distribution Functions (PDFs) and the proton structure functions at small values of \textit{x} (the Bjorken scaling variable, which represents the proton momentum fraction carried by the struck quark) is interesting and the subject of some debates during the last decades~\cite{Heidari:2019fio,Abt:2016vjh,Abdolmaleki:2018jln}. One of the fascinating phenomenological models that well describes experimental data at low \textit{x} is based on the scaling behavior or fractal behavior of PDFs ~\cite{Dremin:1988mu,Lipa:1989yh,Florkowski:1990ba,Bjorken:1994uf,Bhattacharya:2000uh}. Fractal geometry provides a new perspective to describe the natural phenomena and they are based on self-similarity symmetry with respect to scale or size: A small part of the system is approximately similar to the entire system. Benoit Mandelbrot, as a pioneer scientist, defined the fractal as a set of self-similar objects that characterized by the fractional dimension (Hausdorff dimension) which is different than the topological one~\cite{Mandelbrot}. The low \textit{x} behavior of the proton structure is dominated by the gluon-gluon interactions which increase the gluon and the sea quark densities, leading to the self-similarity in PDFs. Recently, relevant work has been done in the investigation of the Deep Inelastic Scattering (DIS) processes~\cite{Dremin:1992zc}. The idea of considering the self-similar features in the multi partons interactions was proposed by Lastovcka~\cite{Lastovicka:2002hw,Lastovicka:2004mq} and others have used it to investigate the proton structure functions ~\cite{Choudhury:2003yy,Choudhury:2005vy,Jahan:2011ig,Choudhury:2013ita,Jahan:2014ova,Jahan:2014sqa,Choudhury:2016fjy,Deppman:2016prl}. In these references, the fractal formalism are applied to proton structure functions up to Leading Order (LO) and they do not consider the gluon densities on their analysis.

 Here in this paper, we intend to use the fractal model to describe the Leading neutron production processes and calculate the neutron Fracture Functions (nFFs) in the positron-proton collisions. These Leading neutrons are high-energy particles that carry a remarkable portion of the incoming proton energy. They are also produced at a small polar angle with respect to the positron-proton collision axis.
The FFs are conditional probabilities of finding partons with flavor \textit{i} and  momentum fraction \textit{x} while simultaneously hadron \textit{h} is detected in the final states of the interaction between lepton and proton. These FFs obey evolution equations with an additional inhomogeneous term as compared to the standard Altarilli-Parisi equations~\cite{deFlorian:1997wi}. They also describe the semi-inclusive DIS processes in the target fragmentation region~\cite{deFlorian:2017lwf,Bertone:2017tyb,Anderle:2015lqa,Ethier:2017zbq,Altarelli:1979kv,deFlorian:1997wi, Daleo:2003jf,Trentadue:1993ka,Trentadue:1994iw}. Although they play a fundamental role in describing the semi-inclusive DIS processes and the global analysis requires their shapes at a given initial scale, QCD is unable to provide or predict them yet. Therefore, it is not only crucial to parametrize them phenomenologically, but also it is necessary to use experimental data in parallel, to make a connection with perturbative QCD.\\
From the experimental point of view, the ZEUS and H1 collaborations published the experimental Leading baryons data~\cite{Adloff:1998yg, Aaron:2010ab, Chekanov:2008tn, Chekanov:2002yh, Chekanov:2002pf, Rinaldi:2006mf}. The semi-inclusive cross-sections for the production of Leading baryons can be studied by using FFs concept~\cite{Shoeibi:2017lrl,Shoeibi:2017zha,Ceccopieri:2014rpa}.\\
In the following, we will investigate the cross-section of  the Leading neutron  production by adopting the fractal model and defining the neutron Fracture Functions, "\textbf{nFFs}". In Sec.~\ref{sec:Theory}, fractals are introduced and the FFs in Leading neutron production are discussed. Furthermore, the emergence of the fractal features in Leading neutron production is examined. In Sec.~\ref{sec:Essential-ingredient}, the parametrization of the neutron FFs are provided within the fractal model and we aso compared them with experimental data. The uncertainties of nFF are given, using the Hessian method. In Sec.~\ref{sec:results}, our findings are checked against H1 experimental data. In Sec.~\ref{sec:Summary}, we finish with the summary and our conclusions.
%
\section{Theory setup}\label{sec:Theory}
In this section, we will review the theoretical aspects of the QCD  analysis performed in this paper. To begin, let us have a short review on the nFFs concept in QCD. They are also reviewed in Refs.~\cite{Ceccopieri:2014rpa,Shoeibi:2017lrl,Shoeibi:2017zha}. We then move to the discussion of the fractal concept and its footprint in the Leading neutron production mechanism.

\subsection{Fracture Functions   and the Leading  neutron Production }\label{subsec:FFs Concept}
Inclusive DIS processes are described by three kinematic variables: $x=\frac{Q^{2}}{2p.q}$, $Q^{2}=-q^{2}$ and $y=\frac{p.q}{p.k}\simeq \frac{Q^{2}}{sx}$ where $p$, $k$, and $q$ are the four-momenta of the incident proton, the incident positron and the exchanged virtual photons, respectively. Here $\sqrt{s}$ is the center of mass energy for positron and proton. However, semi-inclusive DIS describes the mechanism of the Leading neutrons produced in the final state by using two more kinematic variables~\cite{Altarelli:1979kv}: 
\begin{eqnarray}\label{eq:semi-variables}
x_{L}&=&1-\frac{q.(p-p_{n})}{q.p}\simeq \frac{E_{n}}{E_{p}}\nonumber\\
t&=&(p-p_{n})^{2}\simeq -\frac{p_{T}^{2}}{x_{L}}-(1-x_{L})(\frac{m_{n}^{2}}{x_{L}}-m_{p}^{2})
\end{eqnarray}
where $x_{L}$ is the momentum fraction carried by the outgoing neutron and $t$ is squared four-momentum transfer between the incoming proton and the final state neutron. In Eq.~(\ref{eq:semi-variables}), $m_{p}$ is the proton mass and $p_{n}$ is the four-momentum of the final state neutron. The outgoing neutron energy is $E_{n}$ and its transverse momentum for  the incoming proton is denoted by $p_{T}$.

The four-fold differential cross-section for the Leading neutron production, $e^{+}p \rightarrow e^{+}nX $ processes, can be written as a function of semi-inclusive Leading neutron transverse and longitudinal structure functions, $F_2^{\rm Ln(4)}$ and $F_L^{\rm Ln(4)}$, which is defined as~\cite{Aaron:2010ab,Chekanov:2002pf,deFlorian:1998rj, deFlorian:1997wi,Chekanov:2002yh}:
\begin{eqnarray}\label{eq:four-fold}
\frac{d^4\sigma (ep \to e^{\prime} n X)}{d\beta \, dQ^2 \; dx_L \, dt} & = & \frac{4 \pi \alpha^2}{\beta Q^4} (1-y+\frac{y^2}{2}) F_2^{\rm Ln(4)} (\beta, Q^2; x_L, t) \nonumber  \\
& + &  F_L^{\rm Ln(4)} (\beta, Q^2; x_L, t) \,.
\end{eqnarray}
The scaled fractional momentum variable $\beta$ is defined by:
\begin{equation}\label{eq:beta}
\beta=\frac{x}{1-x_L}
\end{equation}
In Eq.~(\ref{eq:four-fold}), $F_{2,L}^{Ln(4)} (\beta, Q^2; x_L,p^2_T)$ can also be interpreted as the sub-structure function of the proton structure functions, $F_{2,L}(x, Q^{2})$ in the  inclusive DIS.
The $t$ integrated differential cross section can be obtained by:
\begin{eqnarray}\label{eq:cross-section2}
\frac{d^3\sigma (ep \to e^{\prime} n X)}{d\beta \, dQ^2 \, dx_L} & = & \int_{t_0}^{t_{min}} \frac{d^4\sigma (ep \to e^{\prime} n X)}{d\beta \, dQ^2 \ dx_L \, dt}  dt \nonumber  \\
& = & \frac{4 \pi \alpha^2}{\beta Q^4} (1-y+\frac{y^2}{2}) F_2^{\rm Ln(3)} (\beta, Q^2; x_L) \nonumber  \\
& + & F_L^{\rm  Ln(3)} (\beta, Q^2; x_L) \,,
\end{eqnarray}
where the integration limits are:
\begin{eqnarray}\label{eq:integration-limits}
t_{min} & = & -(1 - x_L) (\frac{m_N^2}{x_L} - m_p^2)\,, \nonumber  \\
t_0 & = & t_{min} - \frac{(p_T^{max})^2}{x_L} \,.
\end{eqnarray}
$p_T^{max}$ is the upper limit of the neutron transverse momentum used for the $F_{2,L}^{\rm  Ln(3)}$ measurement. In this paper we define the reduced $e^+ p$ cross section $\sigma_r^{\rm  Ln(3)}$ in term of leading neutron transverse $F_2^{\rm  Ln(3)}$ and the longitudinal structure functions $F_L^{\rm  Ln(3)}$ as~\cite{Aaron:2010ab,Chekanov:2002pf}:
\begin{eqnarray}\label{eq:reduced}
\sigma_r^{Ln(3)} & = & F_2^{Ln(3)} (\beta, Q^2; x_L)  \nonumber  \\ 
& - & \frac{y^2}{1+(1-y)^2} F_L^{Ln(3)} (\beta, Q^2; x_L)\,.
\end{eqnarray}

By applying the factorization theorem to the semi-inclusive DIS, we can write the semi-inclusive Leading neutron transverse and longitudinal structure functions, $F_2^{\rm Ln(4)}$ and $F_L^{\rm Ln(4)}$, $F_2^{Ln(4)} (\beta, Q^2; x_L, p^2_T)$ as the convolution of conditional parton distribution functions, $ {\cal M}^n_{i/p} (\beta, \mu_F^2; x_L, p_T^2) $: "\textbf{proton to neutron Fracture Functions}" and Wilson coefficient  functions, \textit{$C_{ki}$} (k = 2, L)~\cite{Ceccopieri:2014rpa}:
\begin{eqnarray}\label{eq:factorization}
F_k^{Ln(4)} (\beta, Q^2; x_L, p_T^2)  = \sum_{i} \int_{\beta}^{1} \frac{d \xi}{\xi} {\cal M}^n_{i/p} (\beta, \mu_F^2; x_L, p_T^2)  \times C_{ki} (\frac{\beta}{\xi}, \frac{Q^2}{\mu_F^2}, \alpha_s(\mu^2_R)) + {\cal O}  (\frac{1}{Q^2}) \,.
\end{eqnarray}

 here $\mu_F^2$ and $\mu^2_R$ are factorization and renormalization scales, respectively and $ {\cal M}^n_{i/p} (\beta, \mu_F^2; x_L, p_T^2) $ describes the probability of detecting Leading neutron with momentum fraction $x_L$ and transverse momentum $p_T^2$ in the final state, while simultaneously the specific parton \textit{i} with fraction momentum of $\beta$ interacts with the virtual photon. Similar to the usual parton densities, these functions are independent of the hard scattering process and represent the objective structure of the target. They also contain rich information about the proton structure functions and non-perturbative dynamics.
Integrating $F_2^{Ln(4)}(\beta, Q^2; x_L, p_T^2)$ with respect to $p_T^2$, up to the small value of $p_T^{max} \ll Q^{2}$, enhances the relative contribution of the exchanged object. Therefore, one may interpret nFFs, ${\cal M}^n_{i/p} (\beta, \mu_F^2; x_L)$, as the parton distribution of the exchanged object (normally pion) between the target and the final state Leading neutron. Using limiting fragmentation hypothesis~\cite{Benecke:1969sh}, $F_2^{Ln(4)} (\beta, Q^2; x_L, p_T^2)$ can be written as the pion flux factor, $f$, (which depends on Leading neutron variable ($x_L, p_T^2$)) and the pion structure function, $F_{2}^{\pi}$, (which depends on the lepton variable ($\beta, Q^2$))~\cite{Chekanov:2002pf}:
\begin{equation}\label{Eq:limiting-fragmentation}
F_2^{Ln(4)} (\beta, Q^2; x_L, p_T^2) = f(x_L, p_T^2)F_{2}^{\pi}(\beta, Q^2)
\end{equation}
Recently different groups~\cite{Arash:2003js,McKenney:2015xis,Barry:2018ort} have used this equation in order to extract the pionic distribution functions by using the experimantal data related to the Leading neutron production~\cite{Chekanov:2002pf,Aaron:2010ab}. It is worth mentioning that for the Leading neutron production process, the pion exchange mechanism is usually considered tobe dominated at large neutron longitudinal momentum fractions $x_L$ and nonpionic contributions become large at small $x_L$. The sensitivity to the one-pion exchange contribution is considered in detail at Ref~\cite{McKenney:2015xis}. 
 The scale dependence of the neutron FFs can be extracted from Leading neutron production processes. Integrating the proton-to-neutron FFs over  $p^2_T$ up to the small value of  $p^{2}_{T_{max}}=\epsilon Q^{2}$ ($\epsilon < 1$), in fixed value of $x_L$, obey the standard DGLAP evolution equations~\cite{Ceccopieri:2014rpa,Ceccopieri:2007th,Shoeibi:2017lrl}:
\begin{equation}\label{eq:DGLAP}
Q^2\frac{\partial M^B_{i/p} (\beta, Q^2; x_L)}{\partial Q^2} = \frac{\alpha_s(Q^2)}{2 \pi} \int_{\beta}^{1} \frac{du}{u} P_{i}^j(u) M^B_{j/p} (\frac{\beta }{u}, Q^2; x_L)
\end{equation}
where $P_{i}^j(u)$ are regularized Altarelli-Parisi splitting functions~\cite{Altarelli:1977zs}. Although the Wilson coefficient functions are  the same as those obtained in inclusive DIS and they are calculable within pQCD~\cite{Vermaseren:2005qc}, the nFFs have non-perturbative nature and their parameterized form should be inspired by and obtained from experimental data.

\subsection{Fractals}\label{subsec:fractals}

\begin{figure}[htb]
	\begin{center}
		\vspace{0.5cm}
		\resizebox{0.9\textwidth}{!}{\includegraphics{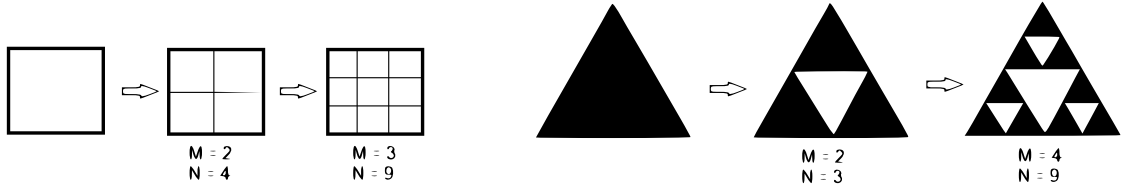}}    
		\caption{ (Color online) Equal segmentation of square and Sierpinski  triangle fractal. \textit{M} is the number of each side segment named it as magnification factor and \textit{N} is the number of self-similar shapes created by the segmentation.}\label{fig:Fractal}
	\end{center}
\end{figure}

Fractal geometry is a new language used to describe the models and analyze the complex forms found in the nature. Plants, clouds,  weather, turbulence,  chaotic phenomena in fluids, biological phenomena such as biological time series or the growth pattern of bacteria and  many other phenomena in nature  can be modeled  by the fractals. Fractals have three general properties: a) Self-similarity. b) Iterative formation. c) Fractional or fractal dimension.
In mathematical language, similarity means that when the size is varied, the form would not be changed. Fig.~(\ref{fig:Fractal}) shows two samples of exact self-similar fractals. By iteration, they are made of smaller and smaller shapes. In this figure, \textit{M} is the number of each side segment named as "magnification factor" and \textit{N} is the number of self-similar shapes created by the segmentation. The self-similarity dimension is introduced in Ref.~\cite{Mandelbrot} and used to consider the proton structure function at low values of \textit{x} in Refs.~\cite{Lastovicka:2002hw,Lastovicka:2004mq}. It defines as:
\begin{equation}\label{Eq:Fractal-Dimention1}
D = \frac{log(number~of~self-similar~objects)}{log(magnification~factor)} = \frac{log~M^{D}}{log~M}
\end{equation}
Using Eq.~(\ref{Eq:Fractal-Dimention1}) for square and Sierpinski triangle fractal in Fig.~(\ref{fig:Fractal}), we achieve the dimension values of these figures as 2 and  1.5849, respectively. The fractional or fractal dimension for the Sierpinski triangle fractal indicates the degree of detail in this object and measures its complexity. The fractal dimension also demonstrates how much space it occupies between the euclidean dimensions. 

Let's look at fractal dimension in more details. Eq.~(\ref{Eq:Fractal-Dimention1}) is the definition of a fractal dimension for discrete fractals. In general, $\zeta$ is a real number and the number of self-similar objects is represented by f($\zeta$). Since the dimension may change with scaling, a local dimension is defined as:
\begin{equation}\label{Eq:Fractal-Dimention2}
D_{f}(\zeta) = \frac{\partial log(f(\zeta))}{\partial log(\zeta)}
\end{equation}

For ideal mathematical fractals, mentioned above, $D_{f}(\zeta)$ is constant for the whole fractal ($D_{f}(\zeta)$$\equiv$D). However, there are many fractals in nature that they are not mathematically ideal and most of the time they have mono fractal structure only in a certain ragion of magnification.as a result, introducing a scale dependent dimension is not weird. If we consider a region on which the the dimension is approximately constant, we can rewrite the Eq.(\ref{Eq:Fractal-Dimention2}) as:

\begin{equation}\label{Eq:Fractal-Dimention}
D = \frac{\partial log~f(\zeta)}{\partial log~\zeta}
\end{equation}
Therefore, the density function can be written as:
\begin{equation}\label{Eq:density-function1}
log~f(\zeta) = D~.~log~\zeta+D_{0}
\end{equation}
In this relation, D$_{0}$ defines the normalization factor, and $f(\zeta)$ is a power-law function ($f(\zeta)\propto \zeta^{D} $) where $D$ is the fractal dimension of this density function. The linear behavior of $logf(\zeta)$ as $log\zeta$ has a key role to find the magnification factor of the density function. For two independent magnification factors, $\zeta$ and $\eta$, this relation can be extended as: 
\begin{equation}\label{Eq:density-function2}
log~f(\zeta, \eta) = D_{\zeta}~.~log~\zeta+D_{\eta}~.~log~\eta+D_{\zeta\eta}~.~log~\eta~log~\zeta+D_{0}
\end{equation}
$D_{\zeta\eta}$ demonstrates the dimensional correlation related to two magnification factors, $\zeta$ and $\eta$. 

\subsection{The Validity test of using the Fractal Model In Leading neutron Production Mechanism}\label{subsec:check-the-validity}

\begin{figure}[htb]
	\begin{center}
		\vspace{0.5cm}
		\resizebox{0.45\textwidth}{!}{\includegraphics{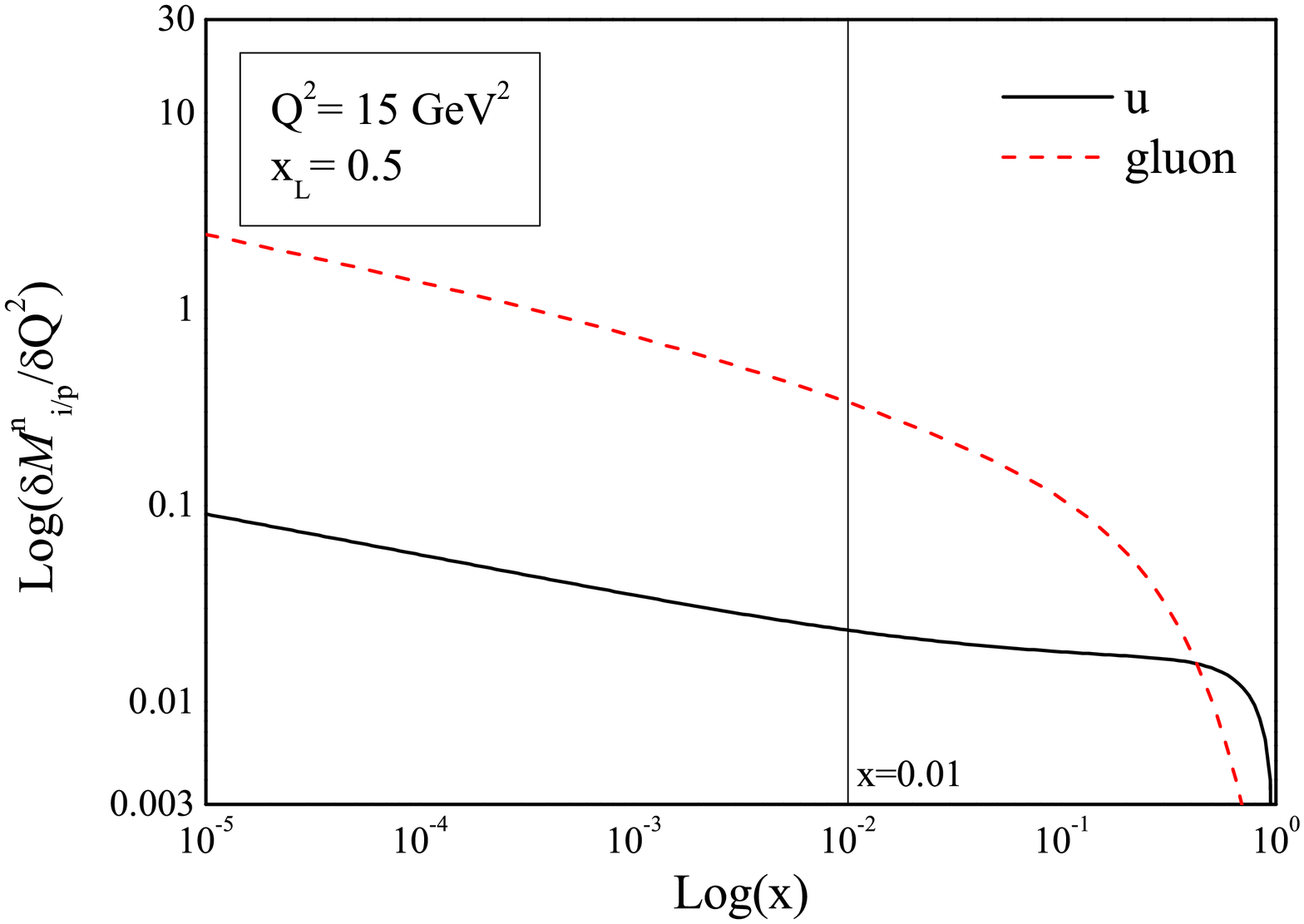}}    
		\resizebox{0.45\textwidth}{!}{\includegraphics{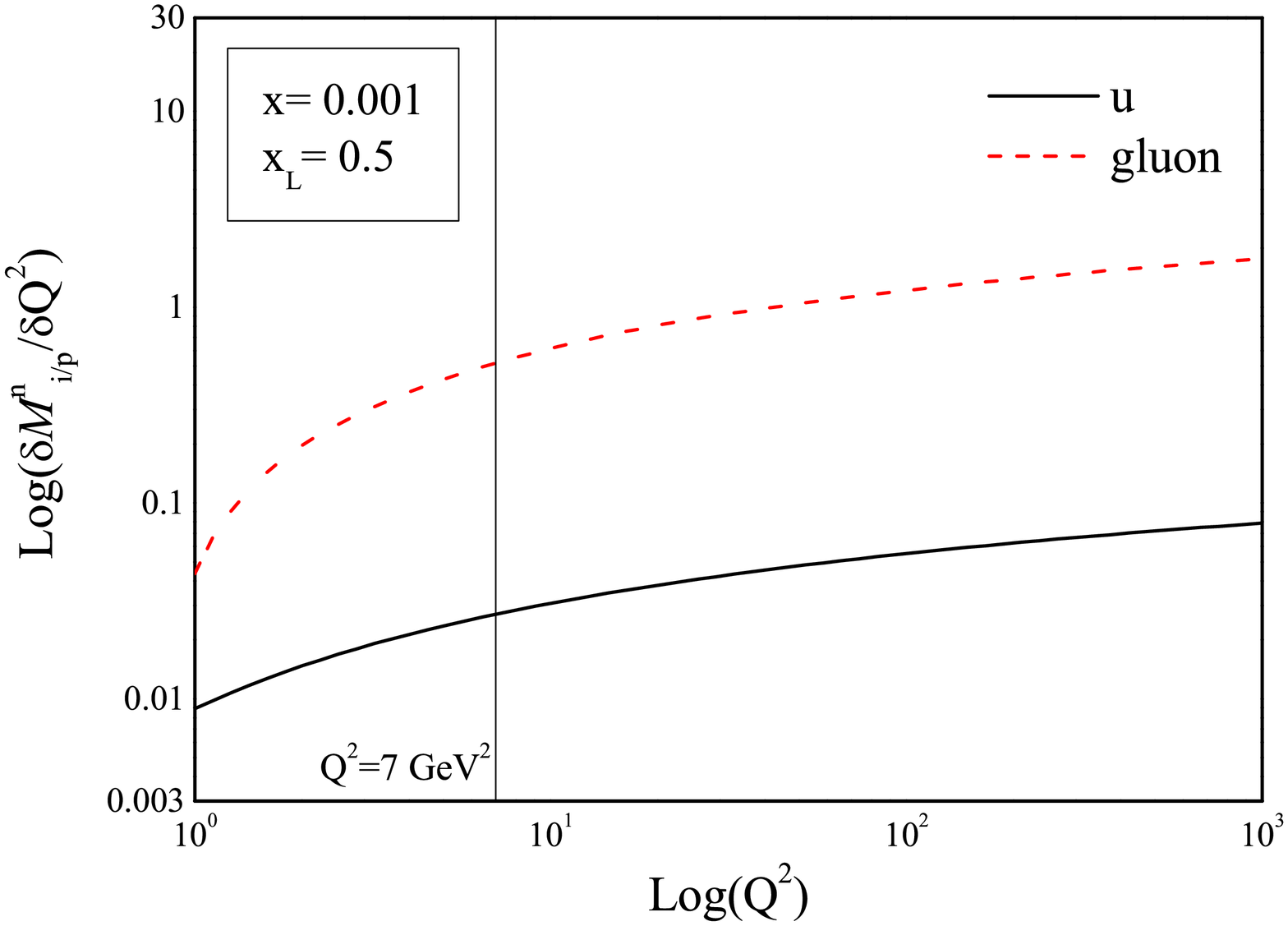}}   
		\caption{ (Color online) The validity  test of the fractal model for Leading neutron production mechanism by using nFFs proposed in Ref~\cite{Shoeibi:2017lrl} }\label{fig:x-Q2-xL}
	\end{center}
\end{figure}

It is known that at low \textit{x} region, scale violation occurs and the parton densities rise. If we assume this region obey the fractal features, the unintegrated parton densities related to the Leading neutron production can be described by the fractal model mentioned in section~\ref{subsec:fractals}. 
In order to check whether the fractal model can be useful in describing the neutron production mechanism or not, we employ the Ref.~\cite{Shoeibi:2017lrl} parameterization of nFFs. This enables us not only to check the validity of the model but also provides us with the magnification factor(s) of the  unintegrated nFFs. The nFFs, ${\cal M}^n_{i/p} (\beta, Q^2; x_L)$ are defined as follows:
\begin{equation}\label{int-unint}
{\cal M}^n_{i/p} (\beta, Q^2; x_L)=  \int_{0}^{Q^{2}}	{\cal M}^n_{i/p} (\beta, k_{t}^2; x_L)~dk_{t}^2
\end{equation}
where $k_t^2$ is the parton transverse momentum. Since unintegrated nFFs, ${\cal M}^n_{i/p} (\beta, k_{t}^2; x_L)$, depending on $\beta=\dfrac{x}{1-x_{L}}$, $x_{L}$ and $k_{t}^{2}$ therefore, it seems that there are three candidates for magnification factors of unintegrated nFFs: $\beta$ (or \textit{x}), $x_{L}$ and $k_t^2$. However, accourding to factorization hypotheses mentioned in Refs.~\cite{H1:1998vul}, the leading neutron transverse structure functions follow the general forms of

\begin{equation}\label{general-form}
F_k^{Ln(3)} (\beta, Q^2; x_L) = f(x_L)~F_{k}^{Ln(2)}(\beta, Q^2)
\end{equation}

where the discrete-function $f(x_L)$ is known as chiral splitting function~\cite{McKenney:2015xis} and it describes the splitting of a proton into a $\pi n$ system. In this analysis this function is expressed as five free parameters.
The parametrized form for F$_{2}^{Ln(2)}(\beta, Q^2)$ is then based on the proton structure functions. Therefore, the two variabe of $\beta$ (or \textit{x}) and $k_t^2$  are the final candidates for magnification factors of unintegrated nFFs. The behaviour of logarithmic unintegrated parton density in order of $log(x)$ in fixed value of $Q^{2}=15~GeV^{2}$ , $x_{L}=0.5$ and its logarithmic behaviour in order of $log(Q^{2})$ in fixed value of $x=0.001$ , $x_{L}=0.5$ are shown in Fig.~(\ref{fig:x-Q2-xL}). The  linear behavior is detected for $x<0.001$ and $Q^2>7 GeV^2$. If we consider the unintegrated densities as fractal, in these two regions, we can find the constant value of fractal dimension by using Eq.~(\ref{Eq:Fractal-Dimention}). This region known as mono fractal region, while in the other regions we expect the multi fractal behavior for those unintegrated PDFs. It means that in Eq.~(\ref{Eq:Fractal-Dimention2}), the fractal dimension should be local one.

As mentioned in Refs.~\cite{Lastovicka:2002hw,Lastovicka:2004mq}, magnification factors have some properties, namely, they should be positive, nonzero and dimensionless. As a result, in order to satisfy the two latter features, we choose $1+\frac{Q^{2}}{q_{0}^{2}}$ instead of $Q^{2}$. In addition,  when the structure is probed deeper, \textit{x} goes to zero while the magnification factor should increase. Therefore, we choose $\frac{1}{x}$ instead of \textit{x}. Consequently, we choose $1+\frac{Q^{2}}{q_{0}^{2}}$ and $\frac{1}{x}$ as magnification factors of unintegrated parton densities in Leading neutron production mechanism. 
\section{Essential parts of the  Fractal Neutron FFs and their uncertainties}\label{sec:Essential-ingredient}
The concept of the Fracture Functions, FFs, is useful to describe and understand the physics in the forward region. Since these universal functions cannot be calculated directly in pQCD, one needs to assume a functional form for them with a number of free parameters and then, attempt to fix the parameters by using the experimental data. Here, we introduce the parameterized form of the nFFs at an initial $Q_0^2$ based on the fractal model. We also  briefly review  the experimental data and estimation of the uncertainties.

\subsection{Neutron FFs in the Fractal approach}\label{subsec:parametrised-nFFs}

Following the sections~\ref{subsec:fractals} and~\ref{subsec:check-the-validity} and Refs.~\cite{Lastovicka:2002hw,Lastovicka:2004mq}, we introduce the unintegrated nFFs as follows:
\begin{equation}\label{Eq:unintegrated-FFs}
log({\cal M}^n_{i/p} (\beta, k_{t}^2; x_L))=\theta(x_L)[D_{1}^{i} log(\frac{1}{\beta})log(1+\frac{{k_{t}^{2}}}{q_{0}^{2}})+D_{2}^{i} log(\frac{1}{\beta})+D_{3}log(1+\frac{{k_{t}^{2}}}{q_{0}^{2}})+D_{0}^{i}-log(M^{2})]
\end{equation}
where q$_{0}$ is a free parameter to be determined in the analysis and  $\theta(x_L)$ is a $x_L$-dependent function. The flavour dependent parameters in Eq.~(\ref{Eq:unintegrated-FFs}) are D$_{0}^{i}$, D$_{1}^{i}$ and D$^{i}_{2}$. D$_{0}^{i}$ fixes the normalization of ${\cal M}^n_{i/p} (\beta, k_{t}^2; x_L)$.  D$_{1}^{i}$ is the dimensional correlation relating to the $\beta$ and Q$^{2}$ factors. D$_{2}^{i}$ and D$_{3}$ are the dimensions related to the magnification factors $\beta$ and Q$^{2}$, respectively. The superscript $\textit{i}$ denotes the gluon and light quarks  (\textit{q}=\textit{u}, $\bar{u}$, \textit{d}, $\bar{d}$, \textit{s}, and $\bar{s}$).
Using the parameterized form given in Eq.~(\ref{Eq:unintegrated-FFs}) and Eq.~(\ref{int-unint}), the initial integrated nFFs (up to Q$_{0}^{2}$ value) are written as:
\begin{eqnarray}\label{eq:input}
\beta{\cal M}^n_{q/p} (\beta, Q_{0}^2; x_L) & = & g(x_{L}).\frac{e^{D_{0}^{q}}q_{0}^{2}\beta^{-D_{2}^{q}+1}}{(M^2)(1+D_{3}-D_{1}^{q}log(\beta))}(\beta^{-D_{1}^{q}log(1+\dfrac{Q_{0}^{2}}{q_{0}^{2}})}(1+\dfrac{Q_{0}^{2}}{q_{0}^{2}})^{D_{3}+1}-1)  \nonumber  \\
\beta{\cal M}^n_{g/p} (\beta, Q_{0}^2; x_L) & = & g(x_{L}).\frac{e^{D_{0}^{g}}q_{0}^{2}\beta^{-D_{2}^{g}+1}}{(M^2)(1+D_{3}-D_{1}^{g}log(\beta))}(\beta^{-D_{1}^{g}log(1+\dfrac{Q_{0}^{2}}{q_{0}^{2}})}(1+\dfrac{Q_{0}^{2}}{q_{0}^{2}})^{D_{3}+1}-1) 
\end{eqnarray}
here $g(x_L)$ is defined as:
\begin{equation}\label{Eq:g(xl)}
g(x_L)= N_{i}x_{L}^{A_{i}}(1-x_{L})^{B_{i}}(1+C_{i}x_{L}^{D_{i}})
\end{equation}
Note that the additional term, -logM$^{2}$, in Eq.~(\ref{Eq:unintegrated-FFs}), having dimension of energy squared, so the integrated parton distribution function defined in Eq. (~\ref{int-unint}) is dimensionless. We set M$^{2}$=1~GeV$^{2}$~\cite{Jahan:2014sqa}.

The power-law behavior of the nFFs can be considered in different point of view: Accourding to Eq.~(\ref{int-unint}), at low value of $\beta$, the unintegrated gluon distribution g($\beta$, $k_{t}^{2}$; $x_L$) is defined as:

\begin{equation}\label{int-unint-g}
xg (\beta, Q^2; x_L)=  \int_{0}^{Q^{2}}	g (\beta, k_{t}^2; x_L)~dk_{t}^2
\end{equation}

g($\beta$, $k_{t}^{2}$; $x_L$) satisfy the BFKL evolution equation:

\begin{equation}\label{BFKL-g}
\dfrac{dg (\beta, k_{t}^2; x_L)}{d ln(1/x)}= K_{L}\otimes g = \lambda g
\end{equation}

where $\otimes$ stands for convolution and $K_{L}$ is the Lipatov kernel which represent the sum over powers of $\alpha_{s}ln(1/x)$ terms~\cite{Lastovicka:2004mq}. From Eq.~(\ref{BFKL-g}) it is obvious that the function \textit{g} follows a power-law behaviour in the variable $\beta$. Therefore, the gluon distribution can be expressed as:

\begin{equation}\label{g-dis}
xg(\beta, Q^2; x_L)\propto f(Q^2) \beta^{-\lambda}
\end{equation}

The special case for which DGLAP equations can be solved analytically is the double leading log approximation (DLL)~\cite{Ninomiya:1976qt}, which is based on the small x limit. The DLL solution represents as:

\begin{equation}\label{g-diss}
g(x, Q^2)~=~e^{2\sqrt{\alpha_{s}^{'}.ln\dfrac{Q^2}{Q_{0}^{2}}. ln\dfrac{1}{x}}}.g(x, Q_{0}^{2})
\end{equation}

here $\alpha_{s}^{'}=\dfrac{3 \alpha_{s}}{\pi}$ and $g(x, Q_{0}^{2})$  is the gluon distribution at factorisation scale Q$_{0}$~\cite{Ninomiya:1976qt}.

In addition, The rise of $F_2(x,Q^2)$ at low \textit{x} can be studied using the following structure function derivative~\cite{H1:2001ert}:

\begin{equation}\label{landa}
\lambda(x, Q^2)=-(\dfrac{\partial ln F_2(x,Q^2)}{\partial ln x})_{Q^{2}}
\end{equation}

Accourding to the H1 publication~\cite{H1:2000muc}, $\lambda(x, Q^2)$ does not depend on \textit{x},
for \textit{x} < 0.01. If we consider the fitting procedure of the form of $F_2(x,Q^2)=c(Q^2)~.~x^{-\lambda(Q^2)}$ to the H1 structure function data~\cite{H1:2000muc}, we see that in the region of x<0.01, there is a nice agreement with the form of $\lambda(Q^2)=a~ln\dfrac{Q^2}{\Lambda^{2}}$ and H1 data. Now Let's come back to our input in Eq.~(\ref{eq:input}).  There are two terms in our proposed model, that they explain nFFs behavior at small value of $\beta$ : $\beta^{-D_{2}^{i}+1}$ and $\beta^{-D_{1}^{i}log(1+\dfrac{Q_{0}^{2}}{q_{0}^{2}})}$. When ones compare the latter term with the above information, they would find out that Eq.~(\ref{eq:input}) is not a weird form for gluon and sea quark densities. It is just like a window that gives us a new perspective on these distribution functions.

As mentioned in Ref.~\cite{Lastovicka:2002hw}, the structure functions of proton is the subsequence of quark density which depend on x and $Q_0^2$ (in this paper $Q_0^2$=$Q^2$). In this reference the Quark-Parton model is used to extract the proton structure function (which depend on \textit{x} and $Q^2$) and the gluon density as an input is not considered. Since nFFs are conditional PDFs, we propose our fractal model by inspiring from Ref.~\cite{Lastovicka:2002hw}. As outlined earlier, nFFs obey the DGLAP evolution equations at a fixed value of $x_L$. Therefore, we set the value of  $Q_{0}^2$=1 GeV$^2$ in Eq.~(\ref{eq:input}) to be able to compare the result of this analysis with the previous one (which are aslo in initial value of $Q_{0}^2$=1 GeV$^2$)~\cite{Ceccopieri:2014rpa,Shoeibi:2017lrl}. Convolution of nFFs with the Wilson coefficients, \textit{$C_i$}, gives the Leading neutron transverse structure functions~\cite{ Ceccopieri:2014rpa,Shoeibi:2017lrl}:
   \begin{eqnarray}\label{eq:F3LN}
 F_k^{Ln(3)} (\beta , Q^2; x_L)  = \sum_{i} \int_{\beta}^{1} \frac{d \xi}{\xi} {\cal M}^n_{i/p} (\beta , \mu_F^2; x_L)  \times C_{ki} (\frac{\beta}{\xi}, \frac{Q^2}{\mu_F^2}, \alpha_s(\mu^2_R)) + {\cal O}  (\frac{1}{Q^2}) \,.
 \end{eqnarray}
Since there are a few experimental data related to the Leading neutron production, in this analysis we assume the symmetric light quark distributions: ${\cal M}^n_{u/p}={\cal M}^n_{\bar{u}/p}={\cal M}^n_{d/p}={\cal M}^n_{\bar{d}/p}={\cal M}^n_{s/p}={\cal M}^n_{\bar{s}/p}$. So, at the initial scale of $Q^{2}_{0}= 1~GeV^{2}$, we have the singlet and gluon distributions as inputs. As fully describe in Ref.~\cite{Ceccopieri:2014rpa}, $g(x_L)$ can be described by common parameters of B$^{g}$=B$^{q}\equiv $B, C$^{g}$=C$^{q}\equiv $C and D$^{g}$=D$^{q}\equiv $D. Therefore the Eq.~\ref{Eq:g(xl)} can be rewritten as: $g(x_L)= N_{i}x_{L}^{A_{i}}(1-x_{L})^{B}(1+Cx_{L}^{D})$. There are two parameters that control the normalization of inputs: N$_{i}$ and D$^{i}_{0}$. We considered these parameters as only one free parameter and we name it as: $\Gamma_{i}$ = N$_{i}e^{D_{0}^{i}}$. As a results, It turns out that the model contains 13 free parameters to be determined using the experimental data ($\Gamma_{i}$, A$_{i}$, B, C, D, q$_{0}$,  D$_{1}^{i}$, D$_{2}^{i}$, and D$_{3}$ ).
To perform the analysis, we used QCDNUM package~\cite{Botje:2010ay} to not only solve the DGLAP evolution equations for the integrated nFFs in Eq.~(\ref{eq:input}), but also we use it to extract the Leading neutron transverse structure functions, $F_2^{Ln(3)} (\beta, Q^2; x_L)$ up to NLO QCD  approximation. We did our analysis in Variable Flavour Number Scheme (VFNS), which means that when the $Q^{2}$ value is equal to the pole mass of the heavy quarks, c, b and t (m$_{c}$=1.4~GeV, m$_{b}$=4.5~GeV and m$_{t}$=175~GeV), the number of flavours change from $n_{f}$ to $n_{f}$+1. In addition, we set the strong coupling constant on it's recently updated average, $\alpha_{s}(M_{z})= 0.1177$~\cite{dEnterria:2015kmd}.

\subsection{Experimental data }\label{subsec:expdata}
The production of Leading neutrons in the
semi-inclusive process $e^{+}$p $\longrightarrow$  $e^{+}$nX are considered by the ZEUS~\cite{Chekanov:2002pf} and H1~\cite{Aaron:2010ab} collaborations at HERA. The H1 experiment measured F$_{2}^{LN(3)}$ in the phase space defined as 6 $<Q^{2}<$100 GeV$^{2}$, 1.5
*10$^{-4}<x<3*10^{-2}$, the longitudinal momentum fraction 0.32$<x_{L}<$0.95 and neutron transverse momentum $p_{T}$<200 MeV. In this interaction, the proton, and positron beam energies are $E_{p}$=920~GeV and $E_{e}$=27.6~GeV, respectively. Notice that the small value of the transverse momentum increases the relative contribution of pion exchange~\cite{Holtmann:1994rs,Kopeliovich:1996iw} and the maximum transverse momentum, $p_{T}^{max}$, is set to 200 MeV~\cite{Adloff:1998yg,Aaron:2010ab}. ZEUS data also cover the large range of kinematics of 7$<Q^{2}<$1000 GeV$^{2}$, 1.1
*10$^{-4}<x<3.2*10^{-2}$ and 0.2$<x_{L}<$1. In this process, positron and proton energies are
$E_{e}$=27.5 GeV and $E_{p}$=820 GeV, respectively and the value of the neutron scattering angle,$\theta_{n}$ , is smaller than 0.8 mrad. The transverse momentum of the neutron is given by $p_{T}\sim x_{L}E_{p}\theta_{n}$.
In this globl analysis, only H1-2010 data are included in fitting procedure.
Fig.~(\ref{fig:x-Q2-xL}) confirms the validity of the fractal model in the kinematic region of \textit{x}$<$0.01 and  Q$^{2}>$7GeV$^{2}$. Therefore, we have limited ourselves to the 189 data points of H1 collaboration, which are compatible with this kinematic region.

\subsection{$\chi^{2}$ minimization and nFFs Uncertainties}

The optimum parameter values of the nFFs at the initial scale of Q$_{0}^{2}$=1~GeV$^{2}$, we used the $\chi^{2}$ function:
\begin{equation}\label{eq:chi2}
\chi^{2}=\sum_{i=1}^{N} \frac{(D_{i}-T_{i})^{2}}{\sigma_{i}^{2}}
\end{equation}
where \textit{i} labels the number of data points used in our  analysis, $D_{i}$ is the data points and $T_{i}$ is the theoretical prediction, which depends on the input nFFs parameters. $\sigma_{i}^{2}$ are the experimental and statistical errors on data points, added in quadrature. In our  analysis, the $\chi^{2}$ function is optimized by the CERN program MINUIT~\cite{James:1994vla}. In table~\ref{tab:fit-parameters}, we give the best fit values of nFFs parameters  at initial scale of Q$_{0}^{2}$=1~GeV$^{2}$.
Those parameters marked by ($^{*}$) are set as fixed values in the final  analysis. In Sec.~\ref{sec:results} we will discuss these obtained parameters in more detail.
In high-energy processes with initial state hadrons, PDFs and reliable knowledge of their uncertainties are needed. PDF uncertainties stem from two sources: a) Theoretical uncertainties which are the consequences of several assumptions made, such as the truncation of the DGLAP perturbation expansion~\cite{Martin:2003sk} b) Experimental uncertainties which are due to the statistical and systematical errors of the experimental data used in the global fit~\cite{Martin:2002aw}. Besides various techniques such as Lagrange Multiplier method~\cite{Stump:2001gu} and Neural Network, the Hessian method is also a reliable method to calculate the experimental uncertainties of PDFs~\cite{Pumplin:2001ct,Martin:2009iq}. We have used it to extract the nFF uncertainties. We refer the reader to Refs~\cite{Pumplin:2001ct,Martin:2009iq} for technical detail. In our NLO QCD analysis, the covariance matrix elements for 9 free parameters are given in  table~\ref{correlation-matrix-elements}.
\begin{table*}[htbp]
	\caption{ The best values, obtained for free parameters in Eq.~(\ref{eq:input}) at the initial scale of Q$_0^2$ = 1 GeV$^2$ using the H1 experimental data. \label{tab:fit-parameters}}
	\begin{tabular}{l|c|cccccccc}
		\hline 
		Parameters & $\beta M^n_{\frac{i=q}{P}} (\beta, Q_0^2, x_L)$  &   $\beta M^n_{\frac{i=g}{P}} (\beta, Q_0^2, x_L)$  \\    \hline
		$\Gamma_{i}$  & $0.14 \pm 0.02$ & $0.61 \pm 0.04$    \\
		$A_{i}$  & $0^* $ & $0^*$     \\
		$B$  & $1.75 \pm 0.02$ & $1.75 \pm 0.02$    \\
		$C$  & $26.3 \pm 1.1$ & $26.3 \pm 1.1$     \\
		$D$  & $6.67 \pm 0.10$ & $6.67 \pm 0.10$     \\
		$q_{0}$  & $0.32 \pm 0.02$ & $0.32 \pm 0.02$    \\
		$D_{1}^{i}$  & $0.033 \pm 0.009$ & $0^* $    \\ 
    	$D_{2}^{i}$ & $1.015 \pm 0.008$ & $1^* $   \\ 
    	$D_{3}$ & $-0.62 \pm 0.02$ & $-0.62 \pm 0.02 $    \\  \hline
	\end{tabular}
\end{table*}

\begin{table*}[htbp]
	\caption{ The correlation matrix elements for the 9 free parameters at NLO fit \label{correlation-matrix-elements}}
	\begin{tabular}{lcccccccccccc}
		\hline 
		 & & $\Gamma_{q}$  & $\Gamma_{g}$  &   $ B$  & $C$& $D$  &   $ q_{0}$   &   $D_{1}^{q} $  & $D_{2}^{q}$ & $D_{3}$ \\     \hline
		$\Gamma_{q}$ & & 1 &  &  & & & & & &   \\
		$\Gamma_{g}$ & & 0.274 & 1 &  &      \\
		$B$  && 0.070 &0.003  & 1 &  & & & & &     \\
		$C$ & & -0.060 & 0.026 & 0.795&  1  & & & & &   \\
		$D$ & & -0.022 & 0.024 &  0.254& 0.534 & 1& & & &   \\
		$q_{0}$ & & -0.503 &-0.806  &  0.127& 0.097  &0.138 &1 & & &   \\
		$D_{1}^{q}$ & & 0.497 & 0.152 & -0.008 & -0.055 &-0.049 &-0.271 &1 & &   \\
		$D_{2}^{q}$ & & -0.206 & 0.063 & -0.019 & 0.014   &0.003 &-0.041 &0.145 &1 &  \\ 
		$D_{3}$ & & -0.340 & -0.583 &0.093 &0.068 &0.099 &0.243 &-0.012 &-0.021 &1   \\  \hline 
	\end{tabular}
\end{table*}

\section{Results and discussion}\label{sec:results}

Let us now give the details of the fractal  analysis. The optimum values obtained in our analysis are summarized in Table~\ref{tab:fit-parameters}. 

\begin{figure}[htb]
	\begin{center}
		\vspace{0.5cm}
		\resizebox{0.45\textwidth}{!}{\includegraphics{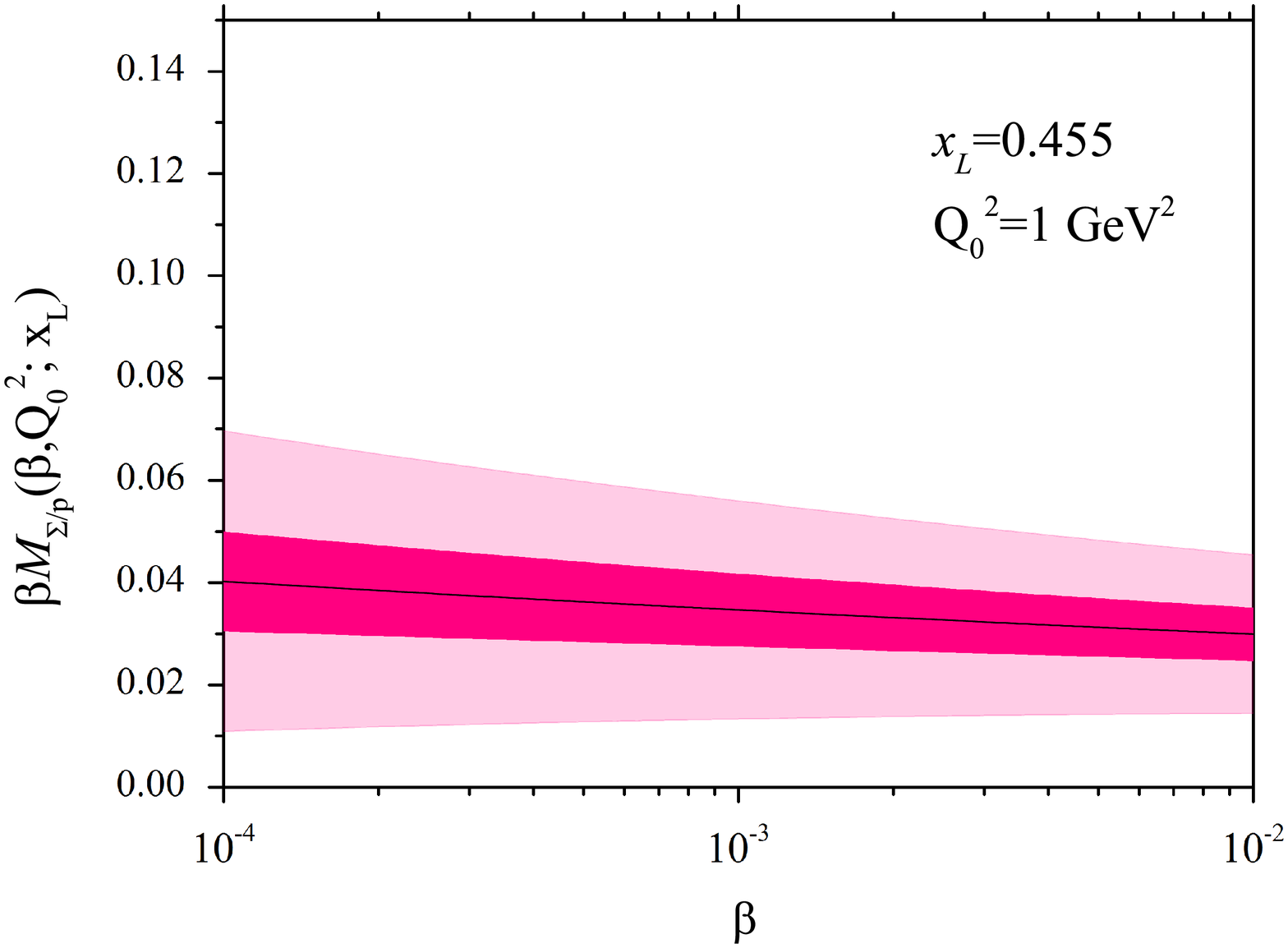}}    
		\resizebox{0.45\textwidth}{!}{\includegraphics{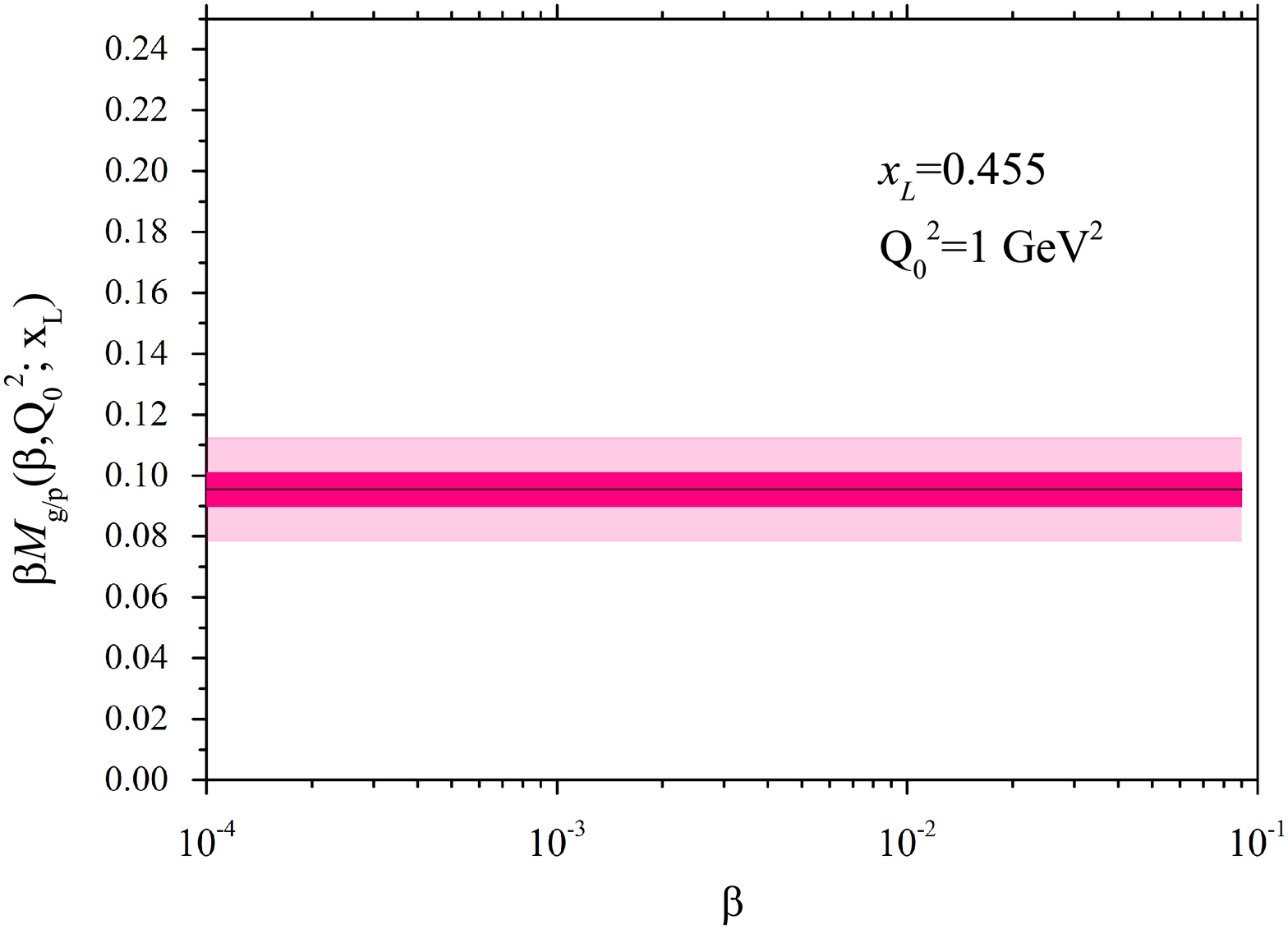}}   
		\resizebox{0.45\textwidth}{!}{\includegraphics{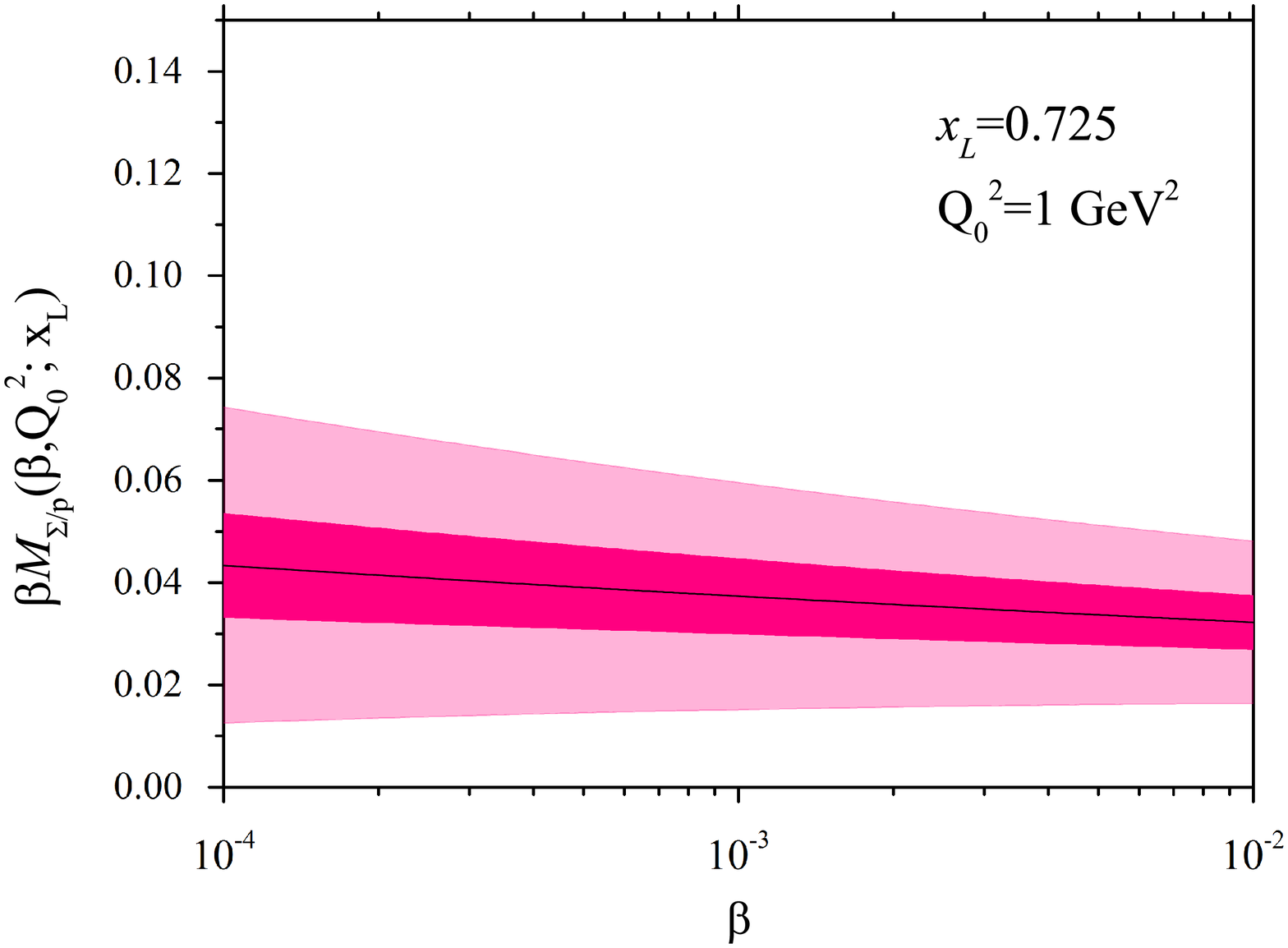}}   
		\resizebox{0.45\textwidth}{!}{\includegraphics{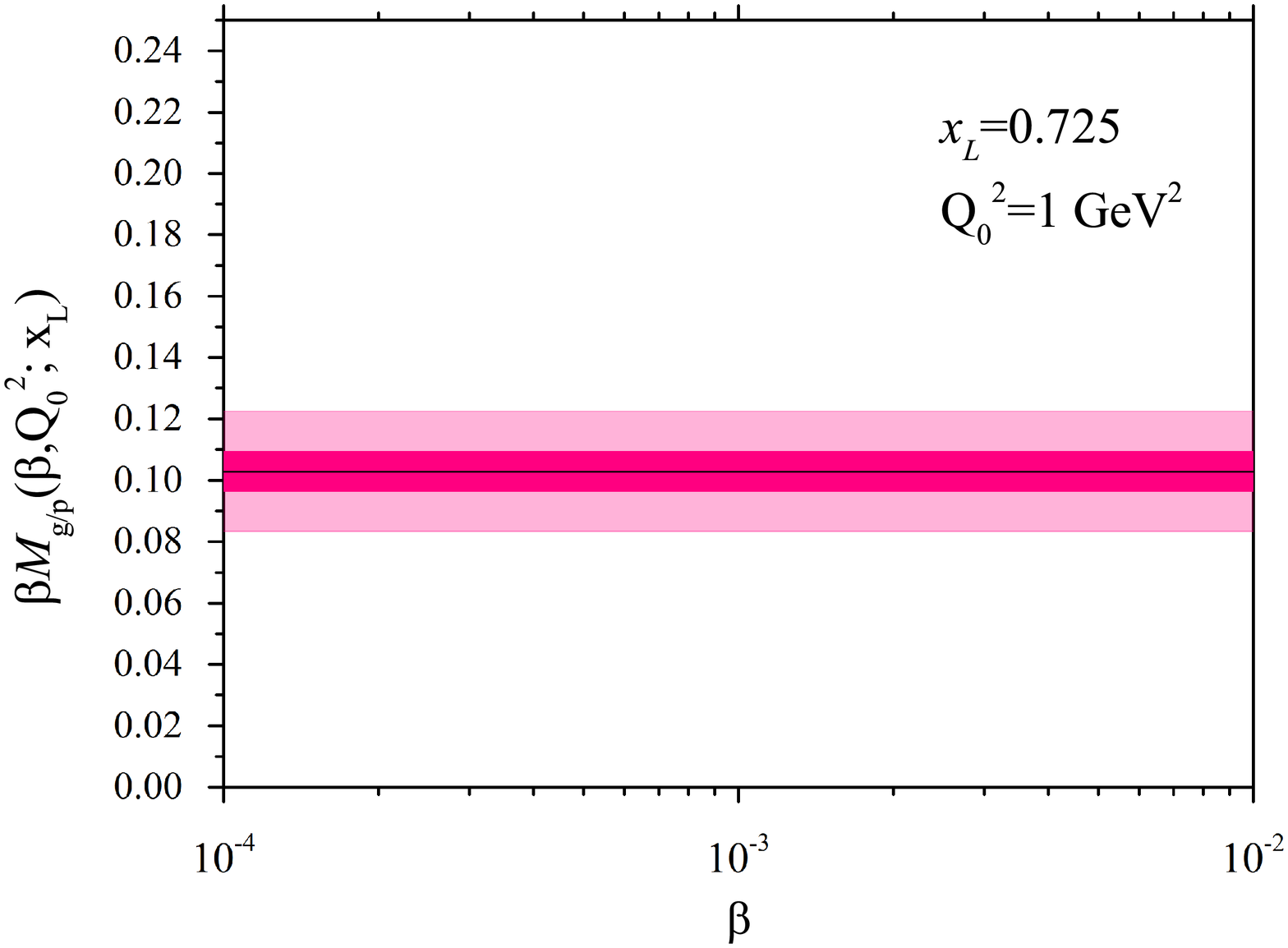}}   
		\caption{(Color online). The singlet and gluon  FFs  as a function of $\beta$ at the initial scale of Q$_{0}^{2}$=1 GeV$^{2}$ for two different values of $x_{L}$=0.455, 0.725.}\label{fig:nFFs}
	\end{center}
\end{figure}

\begin{figure}[htb]
	\begin{center}
		\vspace{0.5cm}
		\resizebox{0.45\textwidth}{!}{\includegraphics{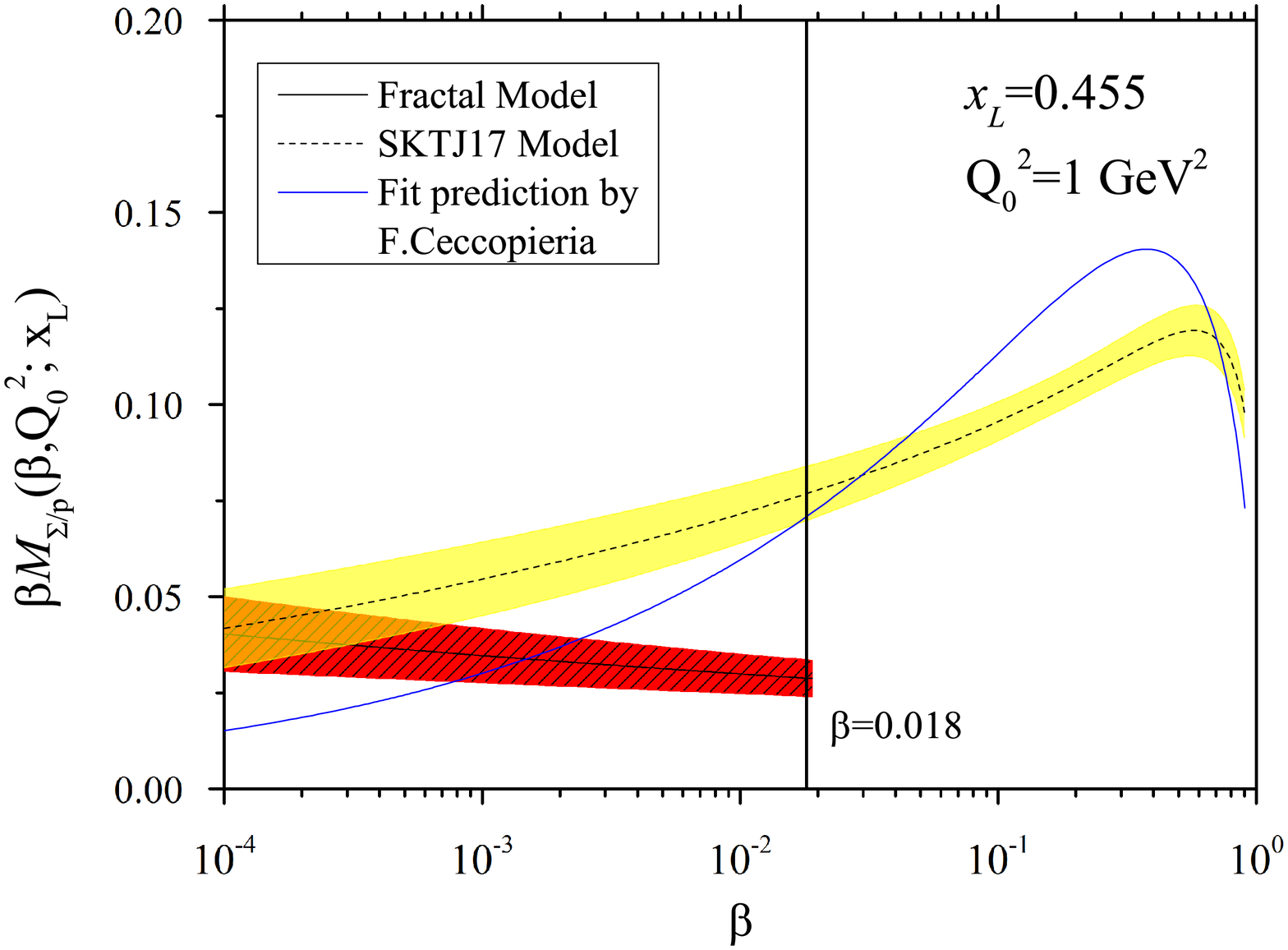}}    
		\resizebox{0.45\textwidth}{!}{\includegraphics{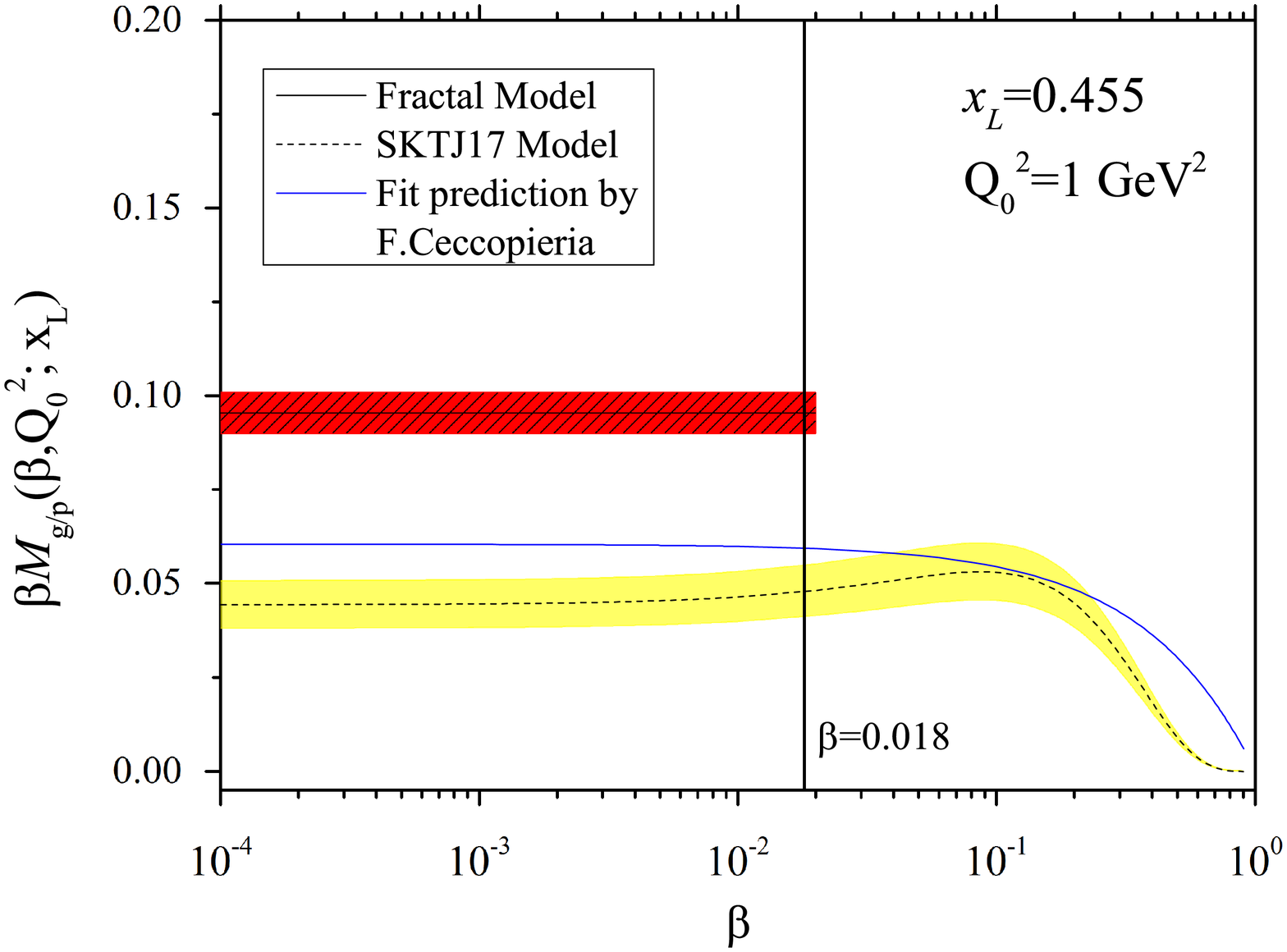}}   
		\resizebox{0.45\textwidth}{!}{\includegraphics{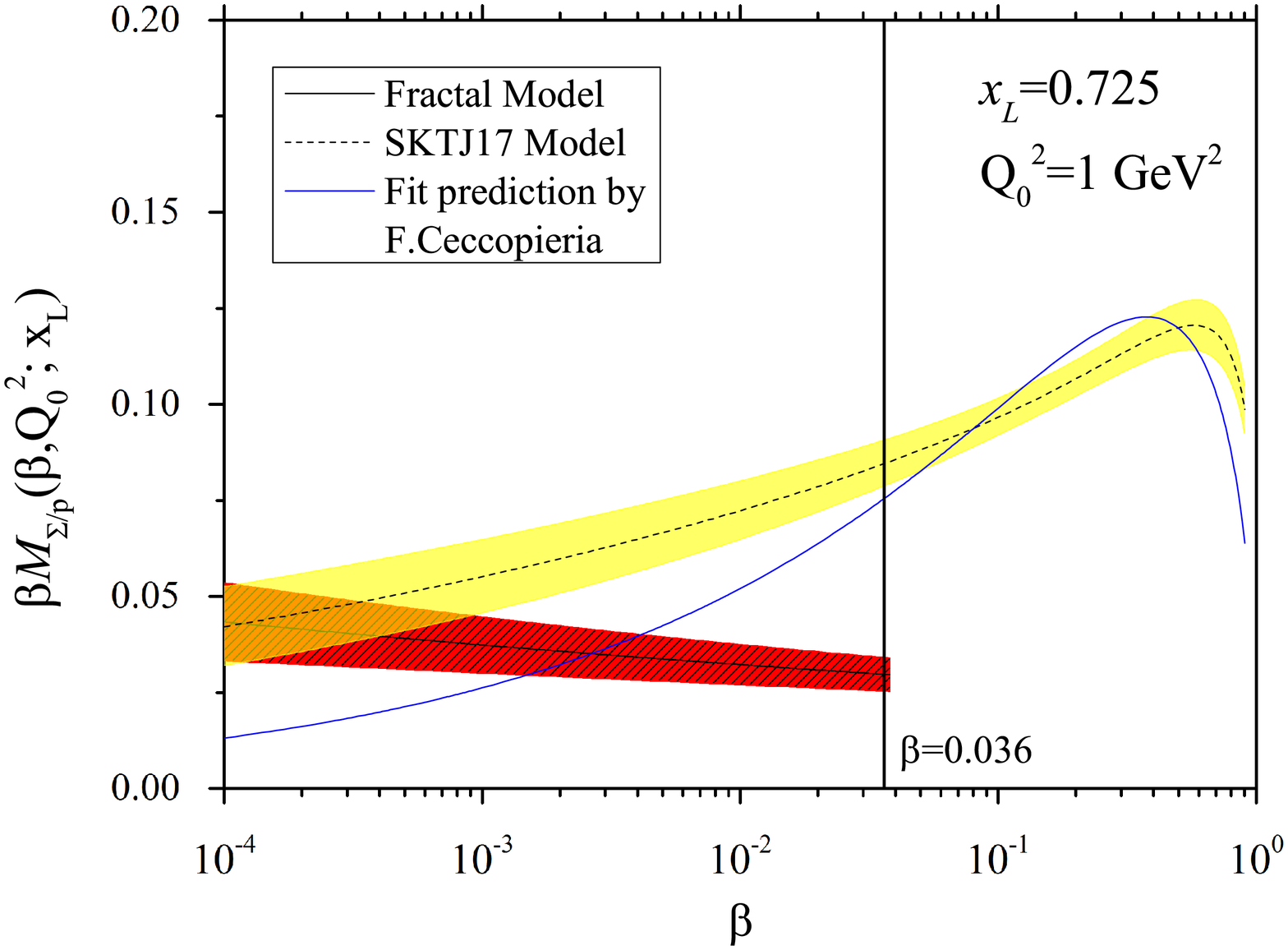}}   
		\resizebox{0.45\textwidth}{!}{\includegraphics{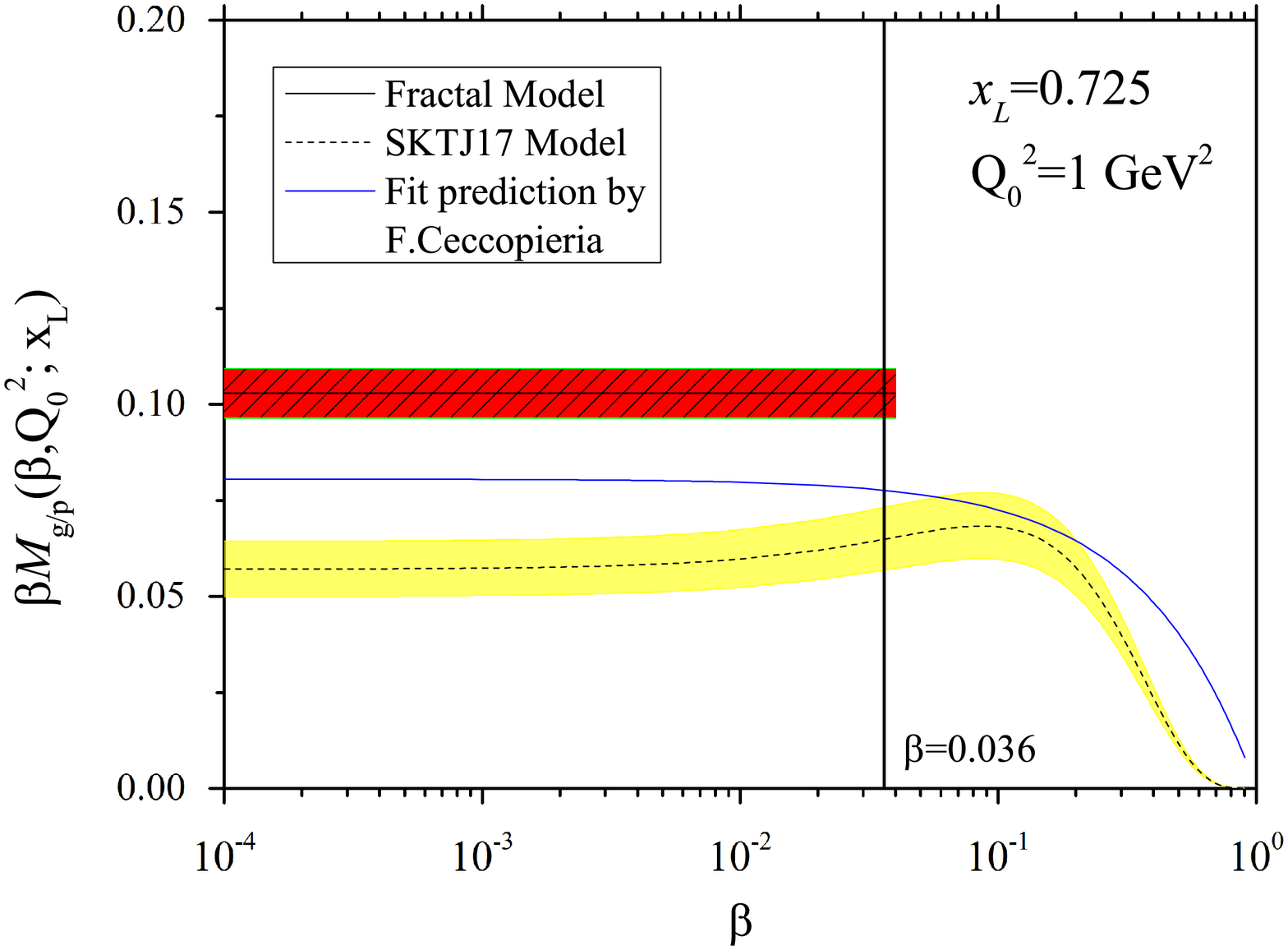}}   
		\caption{ (Color online) Comparing the singlet and gluon momentum distribution resulting from the fractal model with SKTJ17 model~\cite{Shoeibi:2017lrl} and prediction proposed by F.Ceccopieri~\cite{Ceccopieri:2014rpa} }\label{fig:nFFs-compare}
	\end{center}
\end{figure}

\begin{figure}[htb]
	\begin{center}
		\vspace{0.5cm}
		\resizebox{0.45\textwidth}{!}{\includegraphics{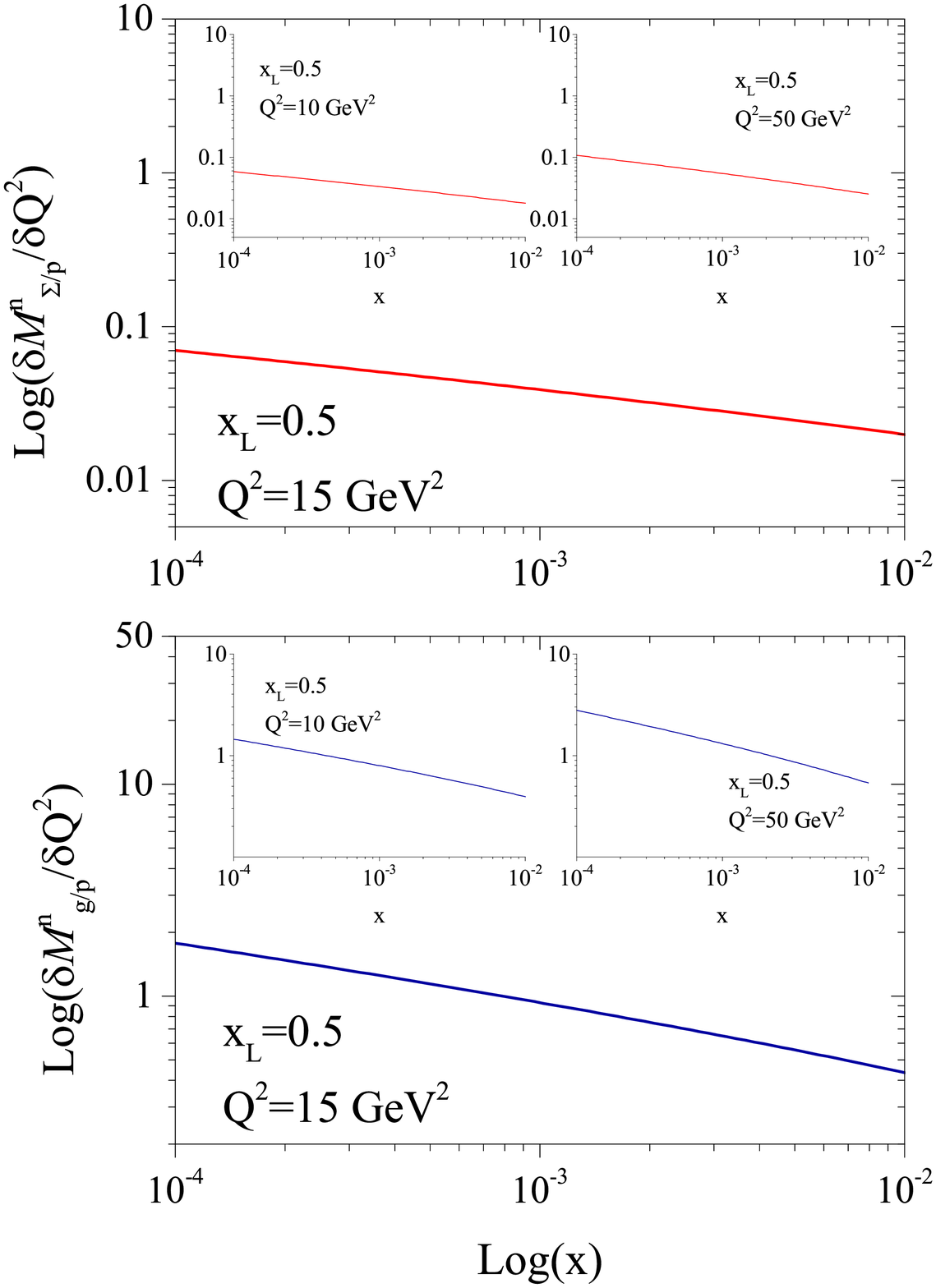}}    
		\resizebox{0.45\textwidth}{!}{\includegraphics{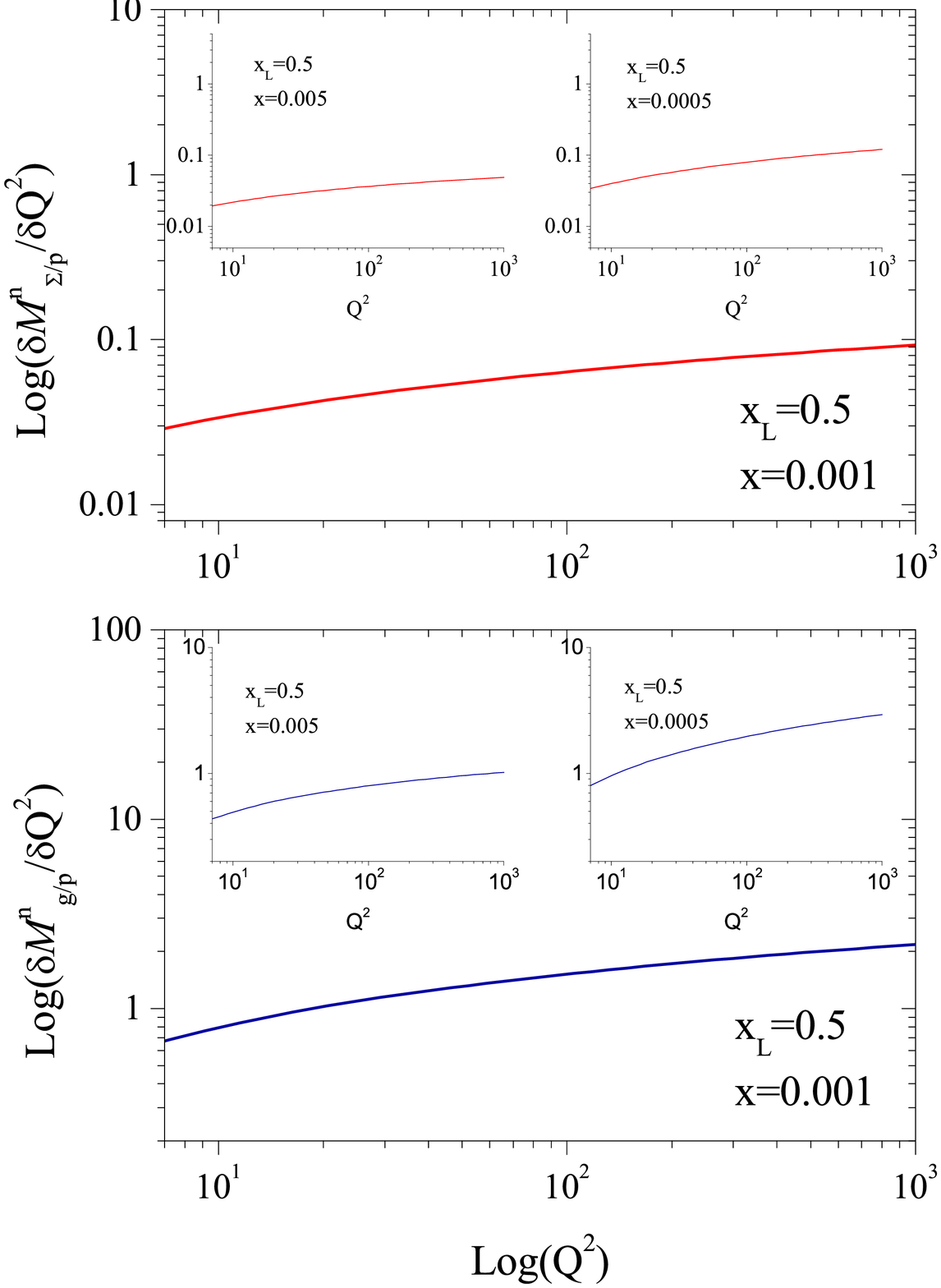}}   
		\caption{ (Color online) The self-similarity behavior of  nFFs in the fixed value of $x_{L}$=0.5 for different values of \textit{x} in some fixed values of Q$^{2}$ (left panel) and also in different values of Q$^{2}$ for some fixed values of \textit{x} (right panel).}\label{fig:nFFs-scaling}
	\end{center}
\end{figure}

\begin{figure}[htb]
	\begin{center}
		\vspace{0.5cm}
		\resizebox{0.7\textwidth}{!}{\includegraphics{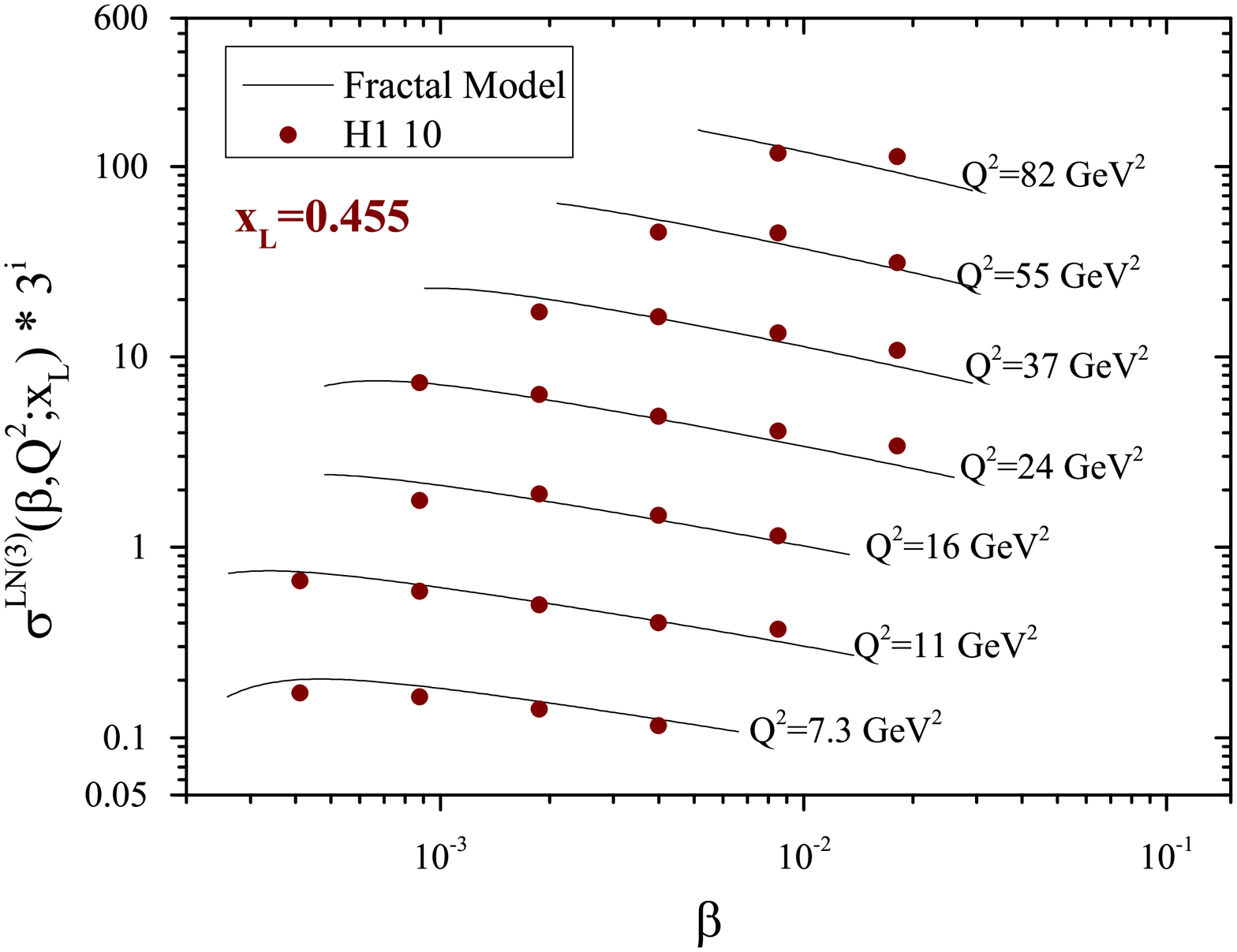}}    
		\resizebox{0.7\textwidth}{!}{\includegraphics{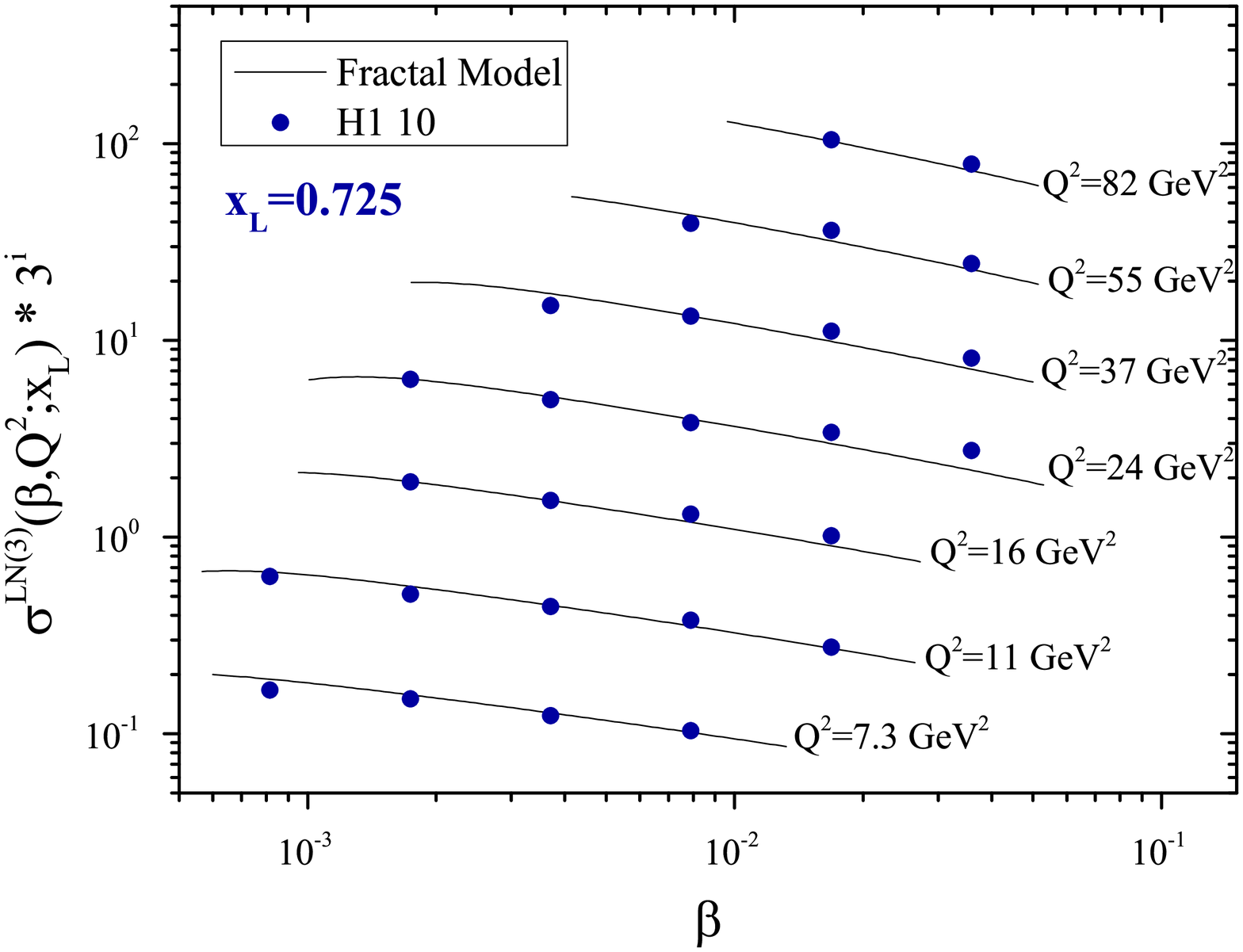}}   
		\caption{ (Color online) The comparison of H1-2010 experimental  data with Leading neutron transverse structure function, $F_{2}(\beta, Q^{2}; x_{L})$,  as a function of $\beta$ for some selected values of Q$^{2}$ at fixed values of x$_{L}$ = 0.455, 0.725. To facilitate the graphical presentation, we have plotted $F_{2}(\beta, Q^{2}; x_{L})$*3$^{i}$. }\label{fig:F2-H1}
	\end{center}
\end{figure}

\begin{figure}[htb]
	\begin{center}
		\vspace{0.5cm}
		\resizebox{0.45\textwidth}{!}{\includegraphics{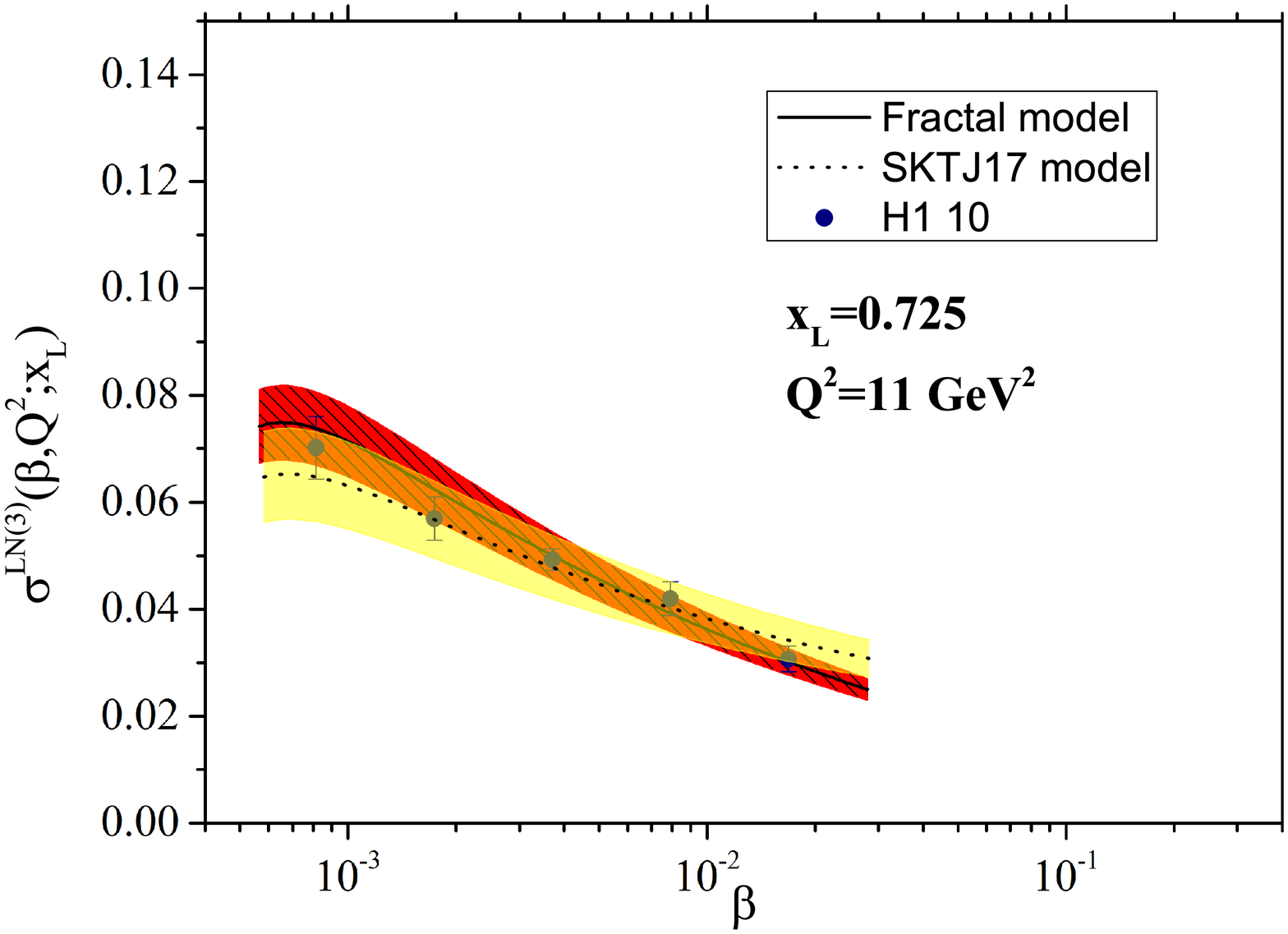}}    
		\resizebox{0.45\textwidth}{!}{\includegraphics{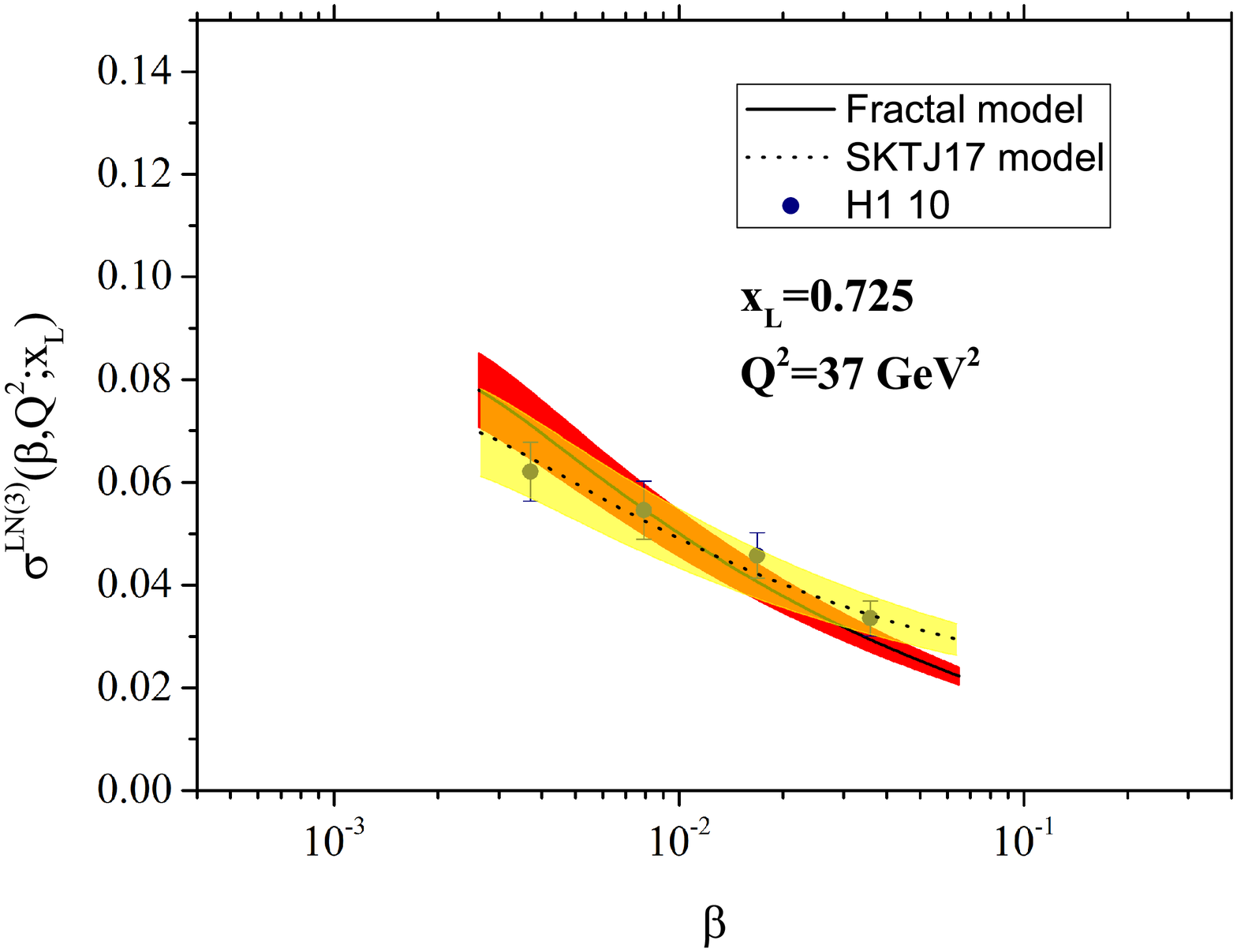}}   
		\resizebox{0.45\textwidth}{!}{\includegraphics{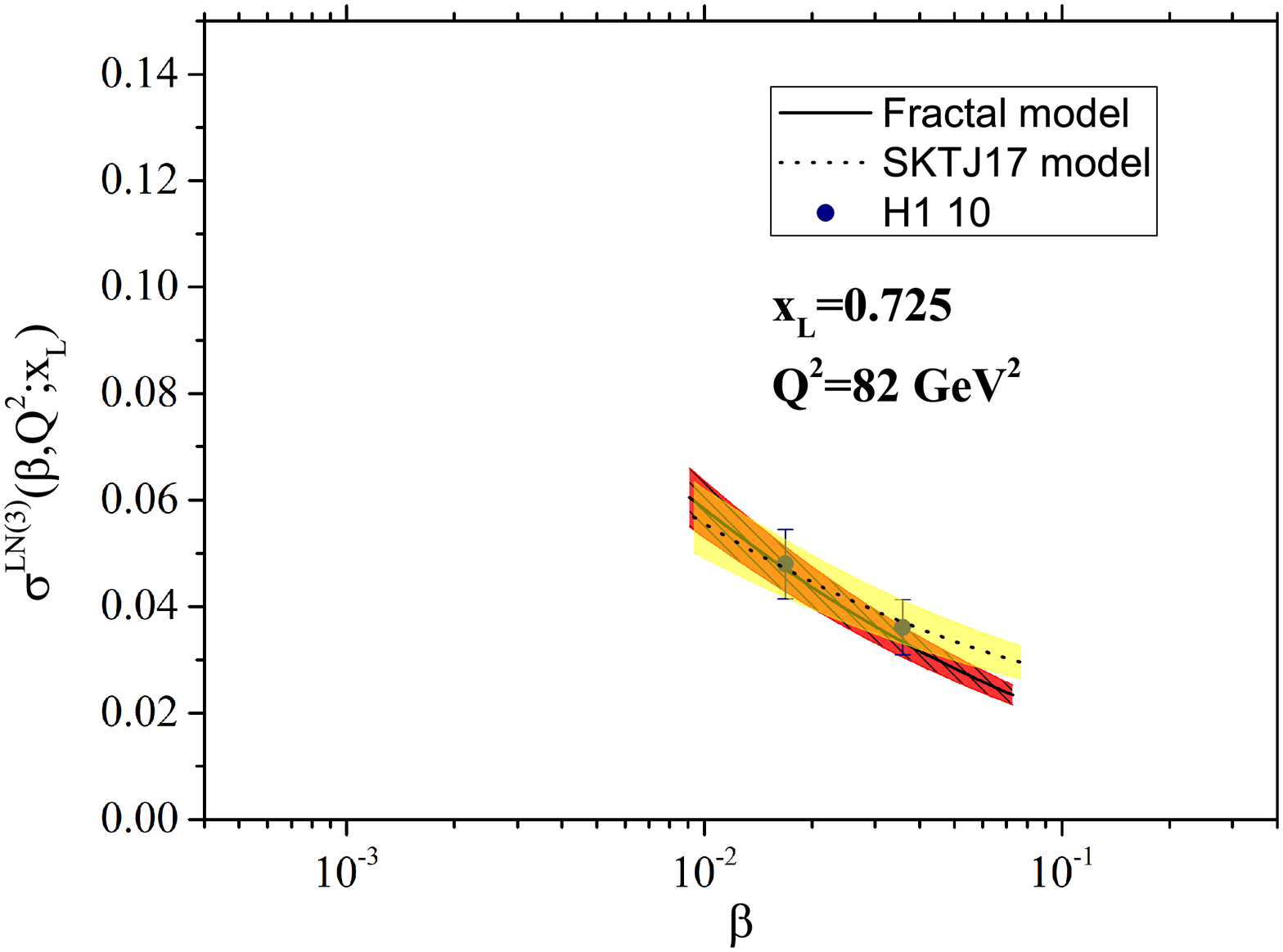}}   
		\caption{ (Color online) The comparison of H1-2010 experimental data with the experimental observables resulting from the fractal model and SKTJ17 model~\cite{Shoeibi:2017lrl}, as a function of $\beta$ for some selected values of Q$^{2}$=11, 37, 82 GeV$^{2}$ at fixed value of x$_{L}$ = 0.725.}\label{fig:compare-F2}
	\end{center}
\end{figure}

In our  analysis, the data set does not include the small values of $x_{L}$, so we choose the zero value for A$_{q}$ and A$_{g}$ parameters. We also obtained a small value with a large error for D$_{1}^{g}$. Therefore, we set its value equal to zero. Now the significant parameter is D$_{2}^{g}$ which controls the small-$\beta$ behavior of gluon FFs. This parameter has so much effect on $\chi^{2}$, so we set its value equal to 1. In this analysis, we obtained  the value of normalized $\chi^{2}$ or the $\chi^{2}$ per degree of freedom as $\frac{\chi^{2}}{n.d.f}=\frac{217.91}{180}=1.21$ which is almost reasonable value.
Our inputs in Eq~(\ref{eq:input}) consist of 13 parameters. In final global analysis we decided to fix 3 parameters. In comparison with the previous analyses~\cite{Ceccopieri:2014rpa,Shoeibi:2017lrl}, the dependence of input to the fixed parameters reduced (in Ref.~\cite{Ceccopieri:2014rpa}, input describe by 11 parameters and 4 parameters are considered to be fixed and in Ref.~\cite{Shoeibi:2017lrl} there is 16 parameters and 8 parameters are chosen to be fixed).
The singlet and gluon densities, $\beta {\cal M}^n_{i/p} (x, Q_{0}^2; x_L)$, resulting from our QCD analysis are shown in Fig.~(\ref{fig:nFFs}) at the initial scale of Q$_{0}^2$=1~GeV$^2$ and at two different values of $x_{L}$=0.455 and 0.725. The shaded bands correspond to the estimated uncertainties of at $\Delta\chi^{2}$=1 and $\Delta\chi^{2}$=10.41  for confidence region of 68\% (see appendix A for more detail).\\
As shown in Fig.~(\ref{fig:nFFs-compare}), our results, based on the fractal model, for $\beta {\cal M}^n_{i/p} (x, Q_{0}^2; x_L)$ is compared with those of SKTJ17 and F. Ceccopieri Models~\cite{Ceccopieri:2014rpa,Shoeibi:2017lrl}. The fractal nFFs seem valid up to the vertical line is shown in the figures,  which confirm the validity of this model at low values of $\beta$. It appears that the upward behavior of the singlet fractal densities at low values of $\beta$ (or \textit{x}) consistent with those we expected from the behavior of the usual PDFs at low \textit{x}.
Fig.~(\ref{fig:nFFs-scaling}) nicely shows the self-similarity behavior of the nFFs in the fixed value of $x_{L}$. The left panel represents the log-log plot of the nFFs for singlet and gluon densities as a function of \textit{x} for different values of Q$^{2}$. The same log-log plot as a function of Q$^{2}$ for different values of \textit{x} is shown in the right panel too.
In Fig.~(\ref{fig:F2-H1}), we present the comparison of our theoretical predictions for Leading neutron transverse structure functions, $F_2^{Ln} (\beta, x_L, Q^2)$, with  H1-2010  experimental data~\cite{Aaron:2010ab}. In general, there is  agreement between our  analysis based on the fractal approach with these experimental data in different values of $\beta$, Q$^{2}$ and $x_{L}$.
Comparison between our result with those from experimental data and SKTJ17 model are shown in Fig.~(\ref{fig:compare-F2}). The shaded bands of both models correspond to the estimated uncertainties of $\Delta\chi^{2}$=1. It seems that the uncertainties in the fractal approach reduce about 50\%.

Finally, as a result of this analysis, we show that the nFFs can be described with a fractal   or self-similar  model at the low values of the fractional momentum $\beta$. Recently,  we used the fractal approach to explore the behavior of un-polarized fragmentation functions (FFs) of the pion in the regions of small momentum fractions \textit{z}~\cite{Mohamaditabar:2020btr}. We believe that it is possible to use the fractal framework to explore the low  \textit{x} behavior of the PDFs too. Work along this line is in progress.
In summary, we show that the upward behavior of the nFFs  for $x<0.001$ and $Q^2>7 GeV^2$ may described by the self-similar behavior of these conditional PDFs. Such behavior has not obtained in the last studies. 

\section{Summary and Conclusions}\label{sec:Summary}

Fractals provide a different way to observe and model complex phenomena. Their applications in strong interaction open detail of high energy collisions such as \textit{ep} scattering.
In 2010 and the years before, several experiments are dedicated to the positron-proton collider at HERA in order to collect precision data for events with the final state related to the Leading neutron production. In this paper, we present the NLO QCD analysis of neutron FFs based on the fractal approach. We have performed an analysis using the H1-2010 Leading neutron production data. As a result, we have shown that the parameterized forms of the neutron FFs based on the fractal concept can give an appropriate description of the Leading neutron production mechanism. We used the Hessian method in order to explain the uncertainties of the nFFs and the corresponding observables.

%
\section*{Acknowledgements}
We gratefully acknowledge to Professor Firooz Arash for carefully reading the manuscript, many helpful discussions, and comments. F. Taghavi Shahri also is thankful to Ferdowsi University of Mashhad for the financial support provided for this research.
This work is supported by the Ferdowsi University of Mashhad under
grant number 73209(25/12/1397).

\newpage\null\thispagestyle{empty}

\begin{appendices}
\section*{Appendix A} 
	
The experimental uncertainties of fractal nFFs are estimated by a general statistical method named as Hessian method. If the optimized parameters of each distribution obtained in this analysis present by $p^{0}_{i}$ (i=1,2,3, ..,N), the quadratic expansion of the $\chi^{2}$  around the minimum point $p^{0}$ is written as:
\begin{equation}\label{Expantion}
\Delta \chi^{2}=\chi^{2}(p^{0}+\delta p)-\chi^{2}(p^{0})=\sum_{i,j}H_{ij} \delta p_{i} \delta p_{j}  \nonumber\\ 
\end{equation}
In this Taylor series expansion, the first derivative terms vanish at the minimum point and $H_{ij}$ is the Hessian matrix defined as
\begin{equation}\label{Hessian}
H_{ij} = \frac{1}{2} \frac{\partial^2 \chi^2} {\partial p_{i} \, \partial p_{j}} \Bigg|_{min}  \nonumber\\
\end{equation}
The uncertainty on a quantity $F(\textit{x},p^{0})$ is then calculated by linear error propagation: 
\begin{equation}\label{Delta-F}
\Delta F = \left[\Delta \chi^2_{\rm global} \, \sum_{i, j=1}^k \frac{\partial F}{\partial p_i} \, H^{-1}_{ij} \, \frac{\partial F}{\partial p_j}\right]^{\frac{1}{2}} \,.  \nonumber\\
\end{equation}
in order to determine the nFFs uncertainties, the confidence region of $\chi^2$ distribution should be estimated.  This region can be identified
by an ellipsoid on this normal distribution and it is defined by $\Delta \chi^{2}$. The confidence level $\textit{P}$  for the $\chi^{2}$ distribution with N
degrees of freedom is written as:
\begin{equation}\label{P6895}
P=\int_{0}^{\Delta \chi^2 }\frac{ (\chi^2)^{\frac{N}{2} - 1}}{ 2^{\frac{N}{2}} \Gamma (\frac{N}{2}) }  e^{\frac{-\chi^2}{2}} d \chi^2 = P (\approx 0.68)~, \nonumber\\
\end{equation}
In this analysis,  we consider  the one-$\sigma$-error range, therefore the confidence level value is $P = 0.68$. Since there are 9 free parameters in the final fitting procedure, by using the above equation, we obtain the value of $\Delta \chi^2=10.41$.
\end{appendices}

\newpage

%

%

\end{document}